\documentclass[a4paper]{ifacconf}

\usepackage{graphicx}      
\usepackage{natbib}          

\usepackage{algorithm}
\usepackage{algorithmicx}
\usepackage{algpseudocode}
\usepackage{amsmath} 
\usepackage{amssymb}  
\usepackage{bm}            
\usepackage{amsfonts}
\usepackage{mathtools}      
\usepackage{mathrsfs}

\usepackage{url}
\usepackage[none]{hyphenat}

\newcommand{\p}{\bm{\mu}}      
\newcommand{\pdim}{n_{\mu}}         
\newcommand{\nx}{n_x}      
\newcommand{\ny}{n_y}
\newcommand{\udim}{n_u}

\newcommand{\msig}{\p_{\sigma}}
\newcommand{\psig}{p_{\sigma}}
\newcommand{\Psig}{P_{\sigma}}

\newcommand{\sigSet}{\sigma \in \Sigma}
\newcommand{\word}{\sigma_{1}\sigma_{2}\cdots \sigma_{k}}
\newcommand{\wordSet}[1]{w \in \Sigma^{#1}}
\newcommand{\expect}[1]{E\left[#1 \right]}
\newcommand{\filt}{\mathcal{F}_{t}^{\p,-}}
\newcommand{\filtp}{\mathcal{F}_{t}^{\p,+}}

\newcommand{\filtr}{\mathcal{F}_{t}^{\mathbf{r}}}

\newcommand{\hankredu}{\mathcal{H}_{ \alpha, \beta }}

\newcommand{\uw}[1]{\p_{#1}}

\newcommand{\zwys}{\mathbf{z}^{\mathbf{y}^{s}}_{w} }
\newcommand{\zwu}{\mathbf{z}^{\mathbf{u}}_{w} }
\newcommand{\zwr}{\mathbf{z}^{\mathbf{r}}_{w} }
\newcommand{\zvr}{\mathbf{z}^{\mathbf{r}}_{v} }

\newcommand{\zwsigs}{\mathbf{z}^{\mathbf{y}^{s}}_{\sigma} }

\newcommand{\timeset}{t \in \mathbb{Z}}
\newcommand{\yb}{\mathbf{y}}
\newcommand{\ybs}{\mathbf{y}^{s}}
\newcommand{\vb}{\mathbf{v}}
\newcommand{\eb}{\mathbf{e}}
\newcommand{\xb}{\mathbf{x}}
\newcommand{\ub}{\mathbf{u}}
\newcommand{\z}{\mathbf{z}}
\renewcommand{\r}{\mathbf{r}}

\newcommand{\varu}{\Lambda_{u}}

\newcommand{\Bs}{G}   

\newcommand{\Tcov}{T_{\sigma,\sigma}}
\newcommand{\covseq}{\Psi_{\mathbf{y}^{s}}}
\newcommand{\incov}{\Psi_{\mathbf{u},\mathbf{y}}}
\newcommand{\alphas}{\bar{\alpha}}
\newcommand{\betas}{\bar{\beta}}

\newcommand{\hankred}{H_{ \alphas, \betas }}

\newtheorem{myprob}{Problem}
\newtheorem{Notation}{Notation}
\newtheorem{Definition}{Definition}
\newtheorem{Example}{Example}

\newtheorem{Corollary}{Corollary}
\newtheorem{Lemma}{Lemma}
\newtheorem{Remark}{Remark}
\newtheorem{Assumption}{Assumption}

\usepackage{enumitem}
\renewcommand{\theenumi}{\arabic{enumi}}

\begin{document}  \sloppy 
\begin{frontmatter}

\title{Realization and identification algorithm for stochastic LPV state-space models with exogenous inputs  \thanksref{footnoteinfo}} 

\thanks[footnoteinfo]{This work was partially funded by CPER Data project, which is co-financed by European Union with the financial support of European Regional Development Fund (ERDF), French State and the French Region of Hauts-de-France.}

\author[First]{Manas Mejari} 
\author[Second]{Mih\'{a}ly Petreczky} 

\address[First]{Centre de Recherche en Informatique, Signal et Automatique de Lille, University of Lille 1, Villeneuve- d'Ascq 59651, France (e-mail: manasdilip.mejari@univ-lille.fr)}
\address[Second]{Centre de Recherche en Informatique, Signal et Automatique de Lille, UMR CNRS 9189, Ecole Centrale de Lille,
	Villeneuve dAscq 59651, France (mihaly.petreczky@ec-lille.fr)}

\begin{abstract}                
In this paper, we present a realization and an identification algorithm for stochastic \emph{Linear Parameter-Varying State-Space Affine} (LPV-SSA) representations. 
The proposed realization algorithm  combines the  deterministic  LPV input output to LPV state-space realization scheme based on correlation analysis with a stochastic covariance  realization  algorithm.
Based on this realization algorithm, a computationally  efficient and statistically consistent identification algorithm is proposed to estimate the LPV model matrices, which are computed from the empirical covariance matrices of outputs, inputs and scheduling signal observations. 
The effectiveness of the proposed  algorithm is shown via a numerical case study.
\end{abstract}

\begin{keyword}
Linear Parameter-Varying systems, stochastic realization.
\end{keyword}

\end{frontmatter}

\section{Introduction}

Identification of \emph{Linear Parameter-Varying} (LPV) models has gained significant attention over the past few years, owing to their ability to describe the behavior of many time-varying and non-linear systems. 
Many approaches have been proposed for the identification of LPV models, in input-output (IO)~\citep{bagi, laurainrefined,MEJARIAuto,pigalpv} as well as state-space (SS)  representations~\citep{fewive,tanelliidentification,vansubspace,veve}.  The reader is referred to~\citep{toth} for a detailed summary of the available LPV identification approaches.

Controller design approaches 
 often require the LPV models to be in SS representation with an affine dependency on the scheduling variable. 
To this end, \emph{realization theory} of LPV models plays a key role in understanding the conditions under which the observed behavior of a system can be realized by a state-space affine representation. It also allows to formulate identification algorithms for estimating state-space representation from a finite set of observations. 
The realization theory for deterministic \emph{Linear Parameter-Varying State-Space with Affine dependence} (LPV-SSA) representation has been developed in~\cite{RolandAbbas,PetreczkyLPVSS}. The results of~\cite{RolandAbbas,PetreczkyLPVSS} were used to derive LPV-SS identification algorithm in \cite{CoxIFAC,CoxLPVSS}. 
These methods are focused mainly on \emph{deterministic} realizations, which for certain control and filtering problems are too restrictive. 

In this paper, we focus on formulating a realization algorithm and a related identification algorithm for stochastic LPV-SSA representations with inputs. The main idea is to decompose the stochastic LPV-SSA realization/identification problem into two independent problems: realization/identification of deterministic part which depends only on the input, 
 and realization/identification of stochastic part.
To this end,  the proposed algorithm is  based on the combination of correlation analysis \citep{CoxLPVSS} for deterministic realization and stochastic covariance identification algorithm for stochastic LPV-SSA representations \citep{MP_CDC19}. 

The  algorithm presented in this paper extends the results of \cite{PetreczkyBilinear,MP_CDC19}, to the case of stochastic LPV-SSA representations with exogenous inputs.  
The  proposed approach  differs significantly from the subspace based identification methods for stochastic LPV-SSA representations \citep{vansubspace,RamosSubspace,FavoreelTAC}.
First, the cited papers do not deal with the realization problem. In particular, while the possibility of decomposing the output into a deterministic and purely stochastic
components is sometimes claimed in the literature, the formal details of such a decomposition were never addressed. 
Second, in contrast to the literature mentioned above, the proposed identification algorithm in this paper
is provenly consistent and it does not require local observability assumptions.
The downside is that the proposed algorithm is provenly consistent only for
a  specific class of scheduling signals and stochastic LPV-SSA representations.
Moreover, the proposed algorithm avoids the curse of dimensionality, but this comes at a price of either using some
prior knowledge on the system to determine the correct selection of the rows and columns of a Hankel-matrix
or using an exhaustive search to find such a  selection. 

The paper is organized as follows. In Section \ref{sec:problem_formulation}, we present the problem formulation.
Section \ref{sect:prop}  presents the formal definition and basic properties of the class of LPV state-space representations
considered in this paper. In Section \ref{sect:decomp}, we formalize the decomposition of outputs of such LPV state-space
representations into stochastic and deterministic components.  In Section \ref{sect:real}, we present the realization algorithm
for stochastic LPV state-space representations, and in Section \ref{sect:ident} we present the related identification algorithm.
Finally, in Section \ref{sec:example} we illustrate the results with a numerical example.

\textbf{Notation}
In the sequel, we will use the standard terminology of probability theory \citep{Bilingsley}. In particular, all 
the random variables and stochastic processes are understood w.r.t. to a fixed probability space $\left(\Omega, \mathcal{F}, \mathcal{P}\right)$, where $\mathcal{F}$ is a $\sigma$-algebra over the sample space $\Omega$ (i.e., $\mathcal{F}$ is a collection of subsets of $\Omega$, that includes $\Omega$ itself, is closed under complement, is closed under countable unions and is closed under countable intersections)
and $\mathcal{P}$ is a probability measure on $\mathcal{F}$. For two $\sigma$-algebras $\mathcal{F}_i$, $i=1,2$, $\mathcal{F}_1 \lor \mathcal{F}_2$ denotes the smallest $\sigma$-algebra generated by the $\sigma$-algebras $\mathcal{F}_1,\mathcal{F}_2$. The expected value of a random variable $\mathbf{x}$ is denoted by $E[\mathbf{x}]$ and conditional expectation w.r.t. $\sigma$- algebra $\mathcal{F}$ is denoted by $\expect{\mathbf{x} \mid \mathcal{F}}$. 
All the stochastic processes in this paper are discrete-time ones defined over the time-axis $\mathbb{Z}$ of the set of integers.
A discrete-time stochastic process is a collection $\{\mathbf{x}(t)\}_{t \in \mathbb{Z}}$ taking values in $X$, where $\mathbf{x}(t) \in X$ is a random variable for all $t \in \mathbb{Z}$.
We denote by $I_n$ the $n \times n$ identity matrix.

\section{PROBLEM FORMULATION}\label{sec:problem_formulation}

Let $\mathbf{y}$, $\mathbf{u}$, $\p$ be stochastic processes taking values in $\mathbb{R}^{\ny}$,
$\mathbb{R}^{\udim}$ and $\mathbb{R}^{\pdim}$ respectively. In this paper, $\mathbf{y}$ represents
the output process, $\mathbf{u}$ is the input process, and
$\p$ is the scheduling signal process.
We define a discrete-time \emph{Linear Parameter-Varying State-Space Affine} (LPV-SSA) representation 
of the process $(\mathbf{y},\mathbf{u},\p)$ as the discrete-time system
of the form
	\begin{align}\label{eqn:LPV_SSA}
	\mathbf{x}(t+1) &= \sum_{i=1}^{\pdim} (A_i\xb(t)+B_i\ub(t)+K_i\vb(t))\p_i(t),
 \nonumber \\ 
	\mathbf{y}(t) &= C\mathbf{x}(t) + D\mathbf{u}(t) + \vb(t),
	\end{align}
where, $A_{i} \in \mathbb{R}^{\nx \times \nx } $, $B_{i} \in \mathbb{R}^{\nx \times \udim } $, $K_{i} \in \mathbb{R}^{\nx \times \ny }$, $\forall i=1, \ldots, \pdim$,  $C \in \mathbb{R}^{\ny \times \nx}$ and $D \in \mathbb{R}^{\ny \times \udim}$ are real  constant matrices, and
$\vb$  is a white noise process, i.e., $E[\vb(t)\vb^{T}(s)]=0$, $s \ne t$ and 
$E[\vb(t)\vb^{T}(t)\p_i(t)]=Q_i > 0$, $i=1,\ldots,\pdim$. 
The realization and identification problems considered in this paper are as follows. 
\begin{myprob}[Realization problem]
\label{problem0}
  For process $(\yb,\ub,\p)$, find matrices $(\{A_{i},B_{i},K_i\}_{i=1}^{\pdim},C,D)$ and processes $\xb,\vb$ such that
  \eqref{eqn:LPV_SSA} is a representation of $(\yb,\ub,\p)$. 
\end{myprob}
\begin{myprob}[Identification problem]
\label{problem1} 
	Assume that $y: \mathbb{Z} \rightarrow \mathbb{R}^{\ny}$ is a sample path of the output process $\yb$, $u: \mathbb{Z} \rightarrow \mathbb{R}^{\udim}$ is a sample path of the input process $\ub$  and $\mu: \mathbb{Z} \rightarrow \mathbb{R}^{\pdim}$ is a sample path of the scheduling process $\p$, corresponding to the same random event $\omega \in \Omega$. 
	Given a dataset $\{{y}(t), u(t), \mu(t) \}_{t=1}^{N}$  consisting of $N$ samples of the output, input and scheduling process, compute from this dataset the estimates $ \{ \{\hat{A}_{i}^N, \hat{B}^N_{i}, \hat{K}^N_{i}, \hat{Q}^N_{i}\}_{i=1}^{\pdim}, \hat{C}^N, \hat{D}^N \}$, such that as $N \rightarrow \infty$, the estimated matrices $ \{ \{\hat{A}_{i}^N, \hat{B}_{i}^N, \hat{K}_{i}^N, \hat{Q}_{i}^N\}_{i=1}^{\pdim}, \hat{C}^N, \hat{D}^N \}$  converge to matrices $ \{ \{A_{i}, B_{i}, K_{i}, Q_{i}\}_{i=1}^{\pdim}, C, D \}$  such that
	the LPV-SSA ~\eqref{eqn:LPV_SSA} with $Q_{i} = E[\vb(t)\vb^{\top}(t) \p^{2}_{i}(t)]$, $i =1,\ldots,\pdim$, is a representation of $(\yb,\ub,\p)$.
\end{myprob}

\section{Properties of LPV-SSA representation}
\label{sect:prop}
In order to make Problems \ref{problem0}-\ref{problem1} well-posed, we have to
 impose additional constraints on the class of processes $(\mathbf{y},\mathbf{u},\p)$ and on the class
 of LPV-SSA representations. 

 Next, we recall from \cite{PetreczkyBilinear} the notion of \emph{Zero Mean Wide Sense Stationary w.r.t. Inputs} (ZMWSSI) process, which will be a central notion for
 the mathematical framework of stochastic LPV-SSA representations. To this end, we need the following
 notation and terminology. 
\begin{Notation}[$\Sigma$]
Let $\Sigma = \{1, \ldots, \pdim\}$. 
\end{Notation}
 The following terminology from automata theory is used.
A \emph{non empty word} over $\Sigma$ is a finite sequence of letters, i.e., $w = \sigma_{1}\sigma_{2}\cdots \sigma_{k}$, where $0 < k \in \mathbb{Z}$, $\sigma_{1}, \sigma_{2}, \ldots, \sigma_{k} \in \Sigma$. The set of \emph{all} nonempty words is denoted by $\Sigma^{+}$. We denote an \emph{empty word} by $\epsilon$. Let $\Sigma^{*} = \epsilon \cup \Sigma^{+}$. 
The concatenation of two nonempty words $v = a_{1}a_{2}\cdots a_{m}$ and  $w= b_{1}b_{2}\cdots b_{n}$ is defined as $vw = a_{1}\cdots a_{m} b_{1} \cdots b_{n}$ for some $m,n > 0$. Note that if $w = \epsilon$ or $v= \epsilon$,  then $v\epsilon = v$ and $\epsilon w = w$, moreover, $\epsilon \epsilon = \epsilon$. The length of the word $w \in \Sigma^{*}$ is denoted by $|w|$, and $|\epsilon| =0$. Example:
for $\pdim=2$,  $\Sigma = \{1,2\}$, $\Sigma^{*}= \{\epsilon,1,2,11,12,21,22,111, \ldots\}$, for the word $w=111 \in \Sigma^{*}$, $|w|=3$.

 \begin{Assumption}[White noise scheduling]
\label{asm:A1}	
The scheduling process $\p = [1, \p_{2}, \ldots, \p_{\pdim}]^{T}$ is zero-mean independent identically distributed (i.i.d.) such that, for all $t \in \mathbb{Z}$, we have $\uw{1}(t) \equiv 1$, and for each $\sigma =2, \ldots, \pdim$, $\uw{\sigma}$ is a zero mean i.i.d. process.
\end{Assumption}
 We define scalars $E[\uw{\sigma}^{2}(t)] = \psig$, for all $t \in \mathbb{Z}$.  In particular, $p_1=1$.

For every word $\wordSet{+}$ where $w=\word$, $k \geq 1$, $\sigma_{1},\ldots, \sigma_{k} \in \Sigma$, we define the  process $\uw{w}$  and
the number $p_w$ as  follows
	\begin{equation*}
        \begin{split}
	 \uw{w}(t) &= \uw{\sigma_{1}}(t-k+1)\uw{\sigma_{2}}(t-k+2)\cdots\uw{\sigma_{k}}(t) , \forall t \in \mathbb{Z} \\
          p_w&=p_{\sigma_1}p_{\sigma_2} \cdots p_{\sigma_k}.
        \end{split}
	\end{equation*}
We set $\uw{\epsilon}(t)=1$ and $p_{\epsilon}=1$. 
For a process $\mathbf{r} \in \mathbb{R}^{\udim}$, for each $\wordSet{+}$ we define the process $\zwr$ as
		\begin{equation}\label{eqn:zwu}
	\zwr(t) = \mathbf{r}(t-|w|) \uw{w}(t-1)\frac{1}{\sqrt{p_{w}}}, \  \forall t \in \mathbb{Z},
		\end{equation}
	which is interpreted as the \emph{past} of $ \mathbf{r}$ w.r.t. $\{\p_{\sigma}\}_{\sigSet}$.
 \begin{Definition}[ZMWSSI, \cite{PetreczkyBilinear}]\label{def:ZMWSSI}
	A stochastic process $\mathbf{r}$ is Zero Mean Wide Sense Stationary w.r.t. the scheduling process
        $\p$ (ZMWSSI) if 
	\begin{enumerate}
		\item For $\timeset$, the $\sigma$-algebras generated by the random variables $\{\mathbf{r}(k) \}_{k \leq t}$, $\{\msig(k) \}_{k < t, \sigSet}$ and $\{\msig(k) \}_{k \geq t, \sigSet},$ denoted by $\filtr$, $\filt$ and $\filtp$ respectively, are such that $\filtr$ and $\filtp$ are conditionally independent w.r.t. $\filt$.
		\item The processes $\{\mathbf{r}, \{\zwr \}_{\wordSet{+}} \}$  are zero mean, square integrable and are jointly wide sense stationary. 
That is, $\forall t,s,k \in \mathbb{Z}$, and for all $w,v \in \Sigma^{+}$, 
		$\expect{\mathbf{r}(t)} = 0$, $\expect{\zwr(t)} =0$, and
	\begin{align*}
	\expect{\r(t+k)(\zwr(s+k))^{T}} &= \expect{\r(t) (\zwr(s))^{T}}, \\
		\expect{\r(t+k)(\r(s+k))^{T}} &= \expect{\r(t)(\r(s))^{T}}, \\
		\expect{\zwr(t+k)(\zvr(s+k))^{T}} &= \expect{\zwr(t) (\zvr(s))^{T}}.
		\end{align*}
\end{enumerate}
\end{Definition}
\begin{Definition}[\cite{PetreczkyBilinear}]
	A process $\r$ is said to be \emph{square integrable w.r.t. $\{\msig\}_{\sigSet}$ (SII process)}, if $\forall \wordSet{*}, \timeset $,
	the random variable $\mathbf{z}^{\r+}_w (t)= \mathbf{r}(t+|w|) \p_{w}(t+|w|-1)\frac{1}{\sqrt{p_w}}$,
is square integrable.
\end{Definition}
 All the process considered in this paper will be assumed to be ZMWSSI and SII process w.r.t. $\p$.  
\begin{Definition}[White noise w.r.t. $\p$]
 A process $\mathbf{r}$ is called a white noise process w.r.t. $\p$, if $\mathbf{r}$ is ZMWSII w.r.t. $\p$,
 and 
 $E[\mathbf{r}(t)(\zwr(t))^T]=0$, $E[\z_{\sigma w}^{\r}(t)(\z_{\sigma w}^{\r}(t))^T]=E[\z^{\r}_{\sigma}(t)(\z^{\r}_{\sigma}(t))^T] > 0$, for all $w \in \Sigma^{+}$, $\sigma \in \Sigma$. 
\end{Definition}
Using the concept of ZMWSSI process and white noise process w.r.t. $\p$, we can formulate the main
assumption regarding the processes $(\mathbf{y},\mathbf{u},\p)$.
\begin{Assumption}
\label{asm:main}
 Assume that $\p$ satisfies Assumption \ref{asm:A1}, and 
 $\begin{bmatrix} \mathbf{y}^T & \mathbf{u}^T \end{bmatrix}^T$ is a ZMWSSI and SII process w.r.t. $\p$, and 
 $\mathbf{u}$ is a white noise process w.r.t. $\p$, and
 the covariance $E[\z^{\ub}_{\sigma}(t)(\z^{\ub}_{\sigma}(t))^{T}]= E[\ub(t-1)(\ub(t-1))^{T}] =  \varu > 0$ does not depend on
 $\sigma \in \Sigma$.  
\end{Assumption}

 Next, we recall from \cite{MP_CDC19} the notion of a \emph{stationary} stochastic LPV-SSA representation of a process $\r$ without inputs.
 \begin{Definition}\label{defn:LPV_SSA_wo_u}
 A \emph{stationary LPV-SSA representation without inputs} of  a process $\r$ taking values in $\mathbb{R}^{p}$, is a tuple $(\{\tilde{A}_{\sigma},\tilde{K}_{\sigma}\}_{\sigma=1}^{\pdim},\tilde{C},\tilde{D}, \tilde{\xb},\tilde{\vb})$, where 
  $\tilde{A}_{\sigma} \in \mathbb{R}^{\tilde{n} \times \tilde{n}}, \tilde{K}_{\sigma} \in \mathbb{R}^{\tilde{n} \times \tilde{m}}$,
  $\tilde{C} \in \mathbb{R}^{p \times \tilde{n}}$ and $\vb$ is a process taking values in $\mathbb{R}^{\tilde{m}}$ such that
  such that 
 \begin{enumerate}
 \item
  $\begin{bmatrix} \tilde{\xb}^T & \tilde{\vb}^T \end{bmatrix}^T$  is a ZMWSSI process, and $E[\z^{\tilde{\xb}}_{\sigma}(t)(\z^{\tilde{\vb}}_{\sigma}(t))^T]=0$,  $E[\tilde{\xb}(t)(\z^{\tilde{\vb}}_w(t))^T]=0$ for
  all $\sigSet$, $ w \in \Sigma^{+}$.
 \item
   $\tilde{\vb}$ is a white noise process w.r.t. $\p$. 
 \item
   The eigenvalues of the matrix $\sum_{\sigSet} \psig \tilde{A}_{\sigma} \otimes \tilde{A}_{\sigma}$ are inside the open unit circle.
 \item
 $\tilde{\xb}(t+1)=\sum_{i=1}^{\pdim} (\tilde{A}_i\tilde{\xb}(t)+\tilde{K}_i\tilde{\vb}(t))\p_i(t)$, $\r(t)=\tilde{C}\tilde{\xb}(t)+\tilde{D}\tilde{\vb}(t)$.
 \end{enumerate} 
  We call $\tilde{\xb}$ the state process and $\tilde{\vb}$ the noise process. 
 \end{Definition}
  In the terminology of \cite{PetreczkyBilinear}, a stationary LPV-SSA without inputs $\ub$, corresponds to a stationary generalized
  bilinear system w.r.t. the scheduling inputs $\{\p_{\sigma}\}_{\sigma \in \Sigma}$. 
  From \cite{PetreczkyBilinear}, if a process $\r$ has a stationary LPV-SSA representation without inputs, then $\r$ is a ZMWSSI process and 
  $\tilde{\xb}$ is uniquely determined by $\tilde{\vb}$ and the matrices $(\tilde{C},\tilde{D}, \{\tilde{A}_{\sigma},\tilde{K}_{\sigma}\}_{\sigma \in \Sigma})$. 
  In order to define this notion more precisely, let us introduce the following notation. 
 \begin{Notation}[Matrix Product]\label{not:product}
Consider a collection of square matrices $A_{\sigma} \in \mathbb{R}^{n \times n}$, $\sigSet$. For any word $\wordSet{+}$ of the form $w = \sigma_{1}\sigma_{2}\cdots\sigma_{k}$, $k\!>\!0$ and $\sigma_{1}, \ldots, \sigma_{k} \in \Sigma$, we define 
	$A_{w} = A_{\sigma_{k}}A_{\sigma_{k-1}}\cdots A_{\sigma_{1}}$.
	For an empty word $\epsilon$, $A_{\epsilon} = I_n$.
\end{Notation}
From \cite{PetreczkyBilinear,MP_CDC19} it follows that
  \begin{equation}
  \label{stat:state:eq1}
  \widetilde{\xb}(t)=\sum_{\sigma \in \Sigma, w \in \Sigma^{*}} \sqrt{p_{\sigma w}}  \tilde{A}_w\tilde{K}_{\sigma}\z^{\tilde{\vb}}_{\sigma w}(t),
 \end{equation}
 where the infinite sum on the right-hand side is  absolutely convergent in the mean square sense.

 Using the notion of a stationary LPV-SSA without inputs, we can define  the class of LPV-SSA representation \emph{with inputs} which 
 will be considered in this paper.
 \begin{Definition}[Stationary LPV-SSA]\label{def:Stationary}
	The LPV-SSA representation \eqref{eqn:LPV_SSA} is \emph{stationary with input $\ub$}, if 
        $(\{A_{\sigma},\begin{bmatrix} K_{\sigma} & B_{\sigma} \end{bmatrix} \}_{\sigma \in \Sigma},C,\begin{bmatrix} I_{\ny} & D \end{bmatrix} \xb,\begin{bmatrix} \vb^T &  \ub^T \end{bmatrix}^T)$ is a stationary LPV-SSA representation of $\yb$ without inputs as in Definition \ref{defn:LPV_SSA_wo_u}, and the orthogonality condition $E[\vb(t)\ub^T(t)\p_{\sigma}^2(t)]=0$, 
     $\forall \sigSet$ holds.
\end{Definition}
  From \eqref{stat:state:eq1} it follows that for a stationary LPV-SSA representation with input $\ub$ of the form \eqref{eqn:LPV_SSA}, 
  \begin{equation*}
     \mathbf{x}(t)\! =  \! \sum_{w \in \Sigma^{*}, \sigma \in \Sigma} \sqrt{p_{\sigma w}} A_w \left(K_{\sigma} \z^{\vb}_{\sigma w}(t)  + B_{\sigma}\z^{\ub}_{\sigma w}(t)\right), 
  \end{equation*} 
  where the infinite sums on the right hand side  are absolutely convergent in the mean-square sense.
  That is, the matrices and the noise processes determine the state process of a stationary LPV-SSA (with or without inputs) uniquely.

\section{Decomposition of the  output of  LPV-SSA representation}
\label{sect:decomp}
It turns out that the output process of  stationary LPV-SSA representations admits a decomposition into 
\emph{deterministic} and \emph{stochastic} parts. The deterministic part depends only on the input process, while the stochastic part depends only on the noise process.
This decomposition does not depend on the particular choice of  LPV-SSA representation, but only on the output process at hand. 

In order to explain this decomposition in more detail, we recall from \cite{PetreczkyBilinear} the
following terminology.
\begin{Notation}[Orthogonal projection $E_l$]
\label{hilbert:notation}
Recall that the set of square integrable
random variables taking values in $\mathbb{R}$, forms a Hilbert-space with the scalar product defined as $<\mathbf{z}_1,\mathbf{z}_2>=E[\mathbf{z}_1\mathbf{z}_2]$. We denote this Hilbert-space by $\mathcal{H}_1$. 
Let $\mathbf{z}$ be a square integrable \emph{vector-valued} 
random variable taking its values in $\mathbb{R}^k$.  Let $M$ be a closed subspace   of $\mathcal{H}_1$. 
By the orthogonal projection of $\mathbf{z}$ onto the subspace $M$, \emph{denoted by $E_l[\mathbf{z} \mid M]$},
we mean the vector-valued square-integrable random variable $\mathbf{z}^{*}=\begin{bmatrix} \mathbf{z}_1^{*},\ldots,\mathbf{z}_k^{*} \end{bmatrix}^T$ such that $\mathbf{z}_i^{*} \in M$ is the orthogonal projection of the $i$th coordinate $\mathbf{z}_i$ of $\mathbf{z}$ onto $M$, as it is usually defined for Hilbert spaces. 
Let $\mathfrak{S}$ be a subset of square integrable random variables in $\mathbb{R}^p$ for some integer $p$, and 
suppose that $M$ is generated by the coordinates of the elements of $\mathfrak{S}$, i.e. $M$ is the smallest (with respect to set inclusion) closed subspace of $\mathcal{H}_1$
which contains the set $\{ \alpha^Ts \mid  s \in \mathfrak{S}, \alpha \in \mathbb{R}^p\}$. 
Then instead of $E_l[z \mid M]$ we will use \( E_{l}[\mathbf{z} \mid \mathfrak{S}] \) to denote the projection of $z$ to $M$.
\end{Notation}
\begin{Definition}[Deterministic and stochastic components]
\label{decomp:outp}
Assume the processes $(\mathbf{y},\mathbf{u},\p)$ satisfy Assumption \ref{asm:main}. Define the \emph{deterministic component} $\yb^d$ of $\yb$ as follows
\begin{equation}
\label{decomp:outp:eq1}
{\yb}^d(t)=E_l[\yb(t) \mid \{\z_w^{\ub}(t)\}_{w \in \Sigma^{+}} \cup \{\ub(t)\}]. 
\end{equation}
Define the \emph{stochastic component} of $\yb$ as 
\begin{equation}
\label{decomp:outp:eq2}
 \yb^s(t)=\yb(t)-\yb^d(t).
\end{equation}
\end{Definition}
 From the definition it follows that
 \[ \yb(t) = {{\yb}}^{d}(t) + {{\yb}}^{s}(t), \]
 i.e., the process $\yb(t)$ can be represented as the sum of its deterministic and  stochastic
 components.  In case when the process admits an LPV-SSA representation, the stochastic and deterministic
 components satisfy the following properties. 
\begin{Lemma}[Decomposition of $\yb$]
\label{decomp:lemma}
 Assume that there exists a stationary LPV-SSA representation of $(\yb,\ub,\p)$ of the form
 \eqref{eqn:LPV_SSA} and that $(\yb,\ub,\p)$ satisfy Assumption \ref{asm:main}. 
  It then follows that
 \begin{align}\label{eqn:LPV_SSA_deter}
{\xb}^{d}(t+1) &= \sum_{i=1}^{\pdim} (A_{i} {\xb}^{d}(t) \!+\! B_{i}  {\ub}(t))\p_i(t),  \nonumber\\ 
{{\yb}}^{d}(t) &= C {\xb}^{d}(t) \!+\! D  {\ub}(t),
\end{align}
and $(\{A_{\sigma},B_{\sigma}\}_{\sigma \in \Sigma},C,D,\xb^d,\ub)$ is a stationary LPV-SSA representation of $\yb^d$ without inputs and with noise process $\ub$, moreover, 
\begin{align}\label{eqn:LPV_SSA_stoch}
{\xb}^{s}(t+1) &= \sum_{i=1}^{\pdim} (A_{i} {\xb}^{s}(t) \!+\! K_{i} {\vb}(t))\p_i(t) ,  \nonumber\\ 
{{\yb}}^{s}(t) &= C{\xb}^{s}(t) \!+\! {\vb}(t),
\end{align}
and $(\{A_{\sigma},K_{\sigma}\}_{\sigma \in \Sigma},C,I_{\ny},\xb^s,\vb)$
is a stationary LPV-SSA representation of $\yb^s$ without inputs, where
\begin{align}
 &\xb^d(t)=E_l[\xb(t)  \mid \{ \z^{\ub}_{w}(t) \}_{w \in \Sigma^{+}} \cup \{\ub(t)\}]
     \label{decomp:lemma:state:eq1} \\
    & \xb^s(t)=\xb(t)-\xb^d(t)
     \label{decomp:lemma:state:eq2} 
\end{align}
\end{Lemma}
The proof of Lemma \ref{decomp:lemma} is presented in \cite[Appendix \ref{app:proof_lem1}]{MP_LPVS19_Arx}.
Thus, $\yb^s$  depends only on the noise $\vb$, and $\yb^d$ does not depend on the noise but it depends
only on input $\ub$. 
In fact, the converse of Lemma \ref{decomp:lemma} also holds. 
\begin{Lemma}
\label{decomp:lemma:inv}
 Assume that $\yb$ has a stationary LPV-SSA representation with input $\ub$. 
 Assume that $\Sigma_d=(\{\hat{A}^{d}_{i}, \hat{B}^{d}_{i} \}_{i=1}^{\pdim}, \hat{C}^{d}, \hat{D}^{d},\hat{\xb}^d,\ub)$ is a 
 stationary LPV-SSA representation of $\yb^d$ without input such that its noise process equals the input process $\ub$.
 Assume that $\Sigma_s=(\{\hat{A}^{s}_{i}, \hat{K}^{s}_{i} \}_{i=1}^{\pdim}, \hat{C}^{s}, I_{\ny},\hat{\xb}^s,\eb^{s})$
 is a stationary LPV-SSA representation of $\yb^s$ without inputs in \emph{forward innovation form}, i.e., assume that the process  $\eb^s$ is the so called \emph{innovation process} of $\yb^s$ as defined in \cite{MP_CDC19,PetreczkyBilinear}:
  \begin{equation} 
 \label{decomp:lemma:innov}
    \eb^s(t)=\yb^s(t)-E_l[\yb^s(t) \mid \{\z^{\yb^s}_w (t)\}_{w \in \Sigma^{+}}] 
  \end{equation}
  Then, tuple $(\{\hat{A}_{i},\hat{K}_{i},\hat{B}_{i}\}_{i=1}^{\pdim},\hat{C},\hat{D},\hat{\xb},\eb^s)$ is a stationary
  LPV-SSA representation of $\yb$ with input $\ub$, where
  \begin{equation} 
 \label{decomp:lemma:sys:eq1}
  \begin{split}
  & \hat{\xb}(t)=\begin{bmatrix} (\hat{\xb}^d(t))^T  & (\hat{\xb}^s(t))^T \end{bmatrix}^T \\ 
  & \hat{A}_{\sigma} = \mathrm{diag}(\hat{A}^{d}_{\sigma},\hat{A}^{s}_{\sigma}), ~ \hat{B}_{\sigma} = \left[ (\hat{B}^{d}_{\sigma})^{T} \ \mathbf{0}_{\nx \times \udim}^{T} \right]^{T} \\
  & \hat{K}_{\sigma} = \left[ \mathbf{0}_{\nx \times \ny}^{T} \  (\hat{K}^{s}_{\sigma})^{T}  \right]^{T}, ~ \hat{C} \!=\! \left[ \hat{C}^{d}  \ \ \hat{C}^{s}  \right], ~ \hat{D} \!=\!  \hat{D}^{d}.
  \end{split}
  \end{equation}
  Moreover, the innovation process $\eb^s$ satisfies
   \begin{equation} 
 \label{decomp:lemma:innov2}
    \eb^s(t)=\yb(t)-E_l[\yb(t) \mid \{\z^{\yb}_w (t), \z^{\ub}_w(t) \}_{w \in \Sigma^{+}} \cup \{\ub(t)\}] 
  \end{equation}
\end{Lemma}
The proof of Lemma \ref{decomp:lemma:inv} is presented in \cite[Appendix \ref{app:proof_lem2}]{MP_LPVS19_Arx}.
Thus, the problem of  realization of $\yb$ can be decomposed into two problems: 
\begin{itemize}
	\item[\textbf{P1}] finding a stationary LPV-SSA representation $\Sigma_d$ without inputs of $\yb^d$, such that the noise process of $\Sigma_d$ is $\ub$, 
	\item[\textbf{P2}]  finding a stationary LPV-SSA representation $\Sigma_s$ without inputs of $\yb^s=\yb-\yb^d$, such that the noise process $\eb^s$ of $\Sigma_s$ is the innovation process of $\yb^s$ as defined in \cite{MP_CDC19,PetreczkyBilinear}. 
\end{itemize}
Moreover, the innovation process $\eb^s(t)$ is the error of projecting $\yb(t)$ onto the linear space spanned by the
 products of the past values of $\yb$, $\ub$ and the scheduling process $\p$, as defined in \eqref{decomp:lemma:innov2}.

In order to solve problem \textbf{P1}, we can adapt realization theory of deterministic LPV-SSA representations.
To this end, in Section \ref{real:alg1} we present an adaptation of the reduced basis Ho-Kalman algorithm from \cite{CoxLPVSS}.
Solution to problem \textbf{P2} was developed in \cite{PetreczkyBilinear}, and a realization algorithm was formulated in
\cite{MP_CDC19}. The latter algorithm will be recalled in Section \ref{real:alg2} which is also based on the reduced basis Ho-Kalman algorithm \citep{CoxLPVSS}.

The combination of realization algorithm from Sections \ref{real:alg1}--\ref{real:alg2} yields a realization algorithm which can 
easily be converted into a system identification algorithm.  
The resulting identification algorithm will first estimate an LPV-SSA representation of $\yb^d$, noise process of which is the input $\ub$, and then it
will estimate a stationary LPV-SSA representation of $\yb^s$ in forward innovation form. 
The identification algorithm outlined above will be presented in Section \ref{sect:ident}.

\section{Realization algorithms}
\label{sect:real}
In this section, we first recall the basis reduced Ho-Kalman realization algorithm for deterministic LPV state-space representations. In turn, this algorithm will be used for covariance realization algorithms for estimating LPV-SSA representations
 of $\yb^d$, $\yb^s$, presented in Section \ref{real:alg1}--\ref{real:alg2}.

\subsection{Basis reduced Ho-Kalman realization algorithm}
\label{sect:real:red}
 
Recall from \cite{PetreczkyLPVSS,CoxLPVSS} that a  deterministic LPV-SSA representation (with affine dependence) is a system of the form
 \begin{equation}
  \label{eqn:LPV_SSA:det} 
  \begin{split}
    x(t+1)&=\sum_{i=1}^{\pdim} (A_ix(t)+B_iu(t))\mu_i(t), ~ \\
    y(t)&=Cx(t)+Du(t),
   \end{split}
 \end{equation}
 where $A_i, B_i, C, D$ are matrices of suitable dimensions,  
$x:\mathbb{Z} \rightarrow \mathbb{R}^{\nx}$ is the state trajectory
$u:\mathbb{Z} \rightarrow \mathbb{R}^{\udim}$ is the input trajectory
$y:\mathbb{Z} \rightarrow \mathbb{R}^{\ny}$ is the output trajectory. In order to avoid technical problems, 
we assume that $x,u,y$ all have finite support, i.e. there exist a $t_0 \in \mathbb{Z}$, such that
$x(s)=0,y(s)=0,u(s)=0$ for all $s < t_0$. We identify a deterministic  LPV-SSA of the form \eqref{eqn:LPV_SSA:det}
with the tuple $\mathscr{S}=(\{A_{\sigma},B_{\sigma}\}_{\sigma \in \Sigma},C,D)$. The number $\nx$ is called the dimension of $\mathscr{S}$. 
The \emph{sub-Markov parameters of $\mathscr{S}=(\{A_{\sigma},B_{\sigma}\}_{\sigma \in \Sigma},C,D)$} are the values of the map $M_{\mathscr{S}}:\Sigma^{*} \rightarrow \mathbb{R}^{\ny \times \udim}$, such
that for all $w \in \Sigma^{*}$, 
\begin{equation}\label{eqn:sub_markov}
M_{\mathscr{S}}(w)=\left\{\begin{array}{ll}
CA_sB_{\sigma}, \ \ & w=\sigma s, \sigma \in \Sigma, s \in \Sigma^{*} \\
D.           \ \  & w=\epsilon
\end{array}\right.
\end{equation}
We will refer to $M_{\mathscr{S}}$ as \emph{the sub-Markov function  of the deterministic LPV-SSA representation of $\mathscr{S}$}.
From \cite{PetreczkyLPVSS} it then follows that two deterministic LPV-SSA representations $\mathscr{S}_1$, $\mathscr{S}_2$ have the same input-output behavior, if and only if
their sub-Markov parameters are equal, i.e., $M_{\mathscr{S}_1}=M_{\mathscr{S}_2}$. Moreover, the sub-Markov parameters can be
determined from the input-output behavior. 

Below we recall from \cite{CoxLPVSS} an adaptation of this Ho-Kalman-like algorithm, which uses sub-Markov parameters to
compute a deterministic LPV-SSA representation.
In order to present the algorithm, we present the notion of \emph{$n$-selection}. 
Let us define the set $\Sigma^{n}$ as the set of all words $w \in \Sigma^{*}$ of length less than or equal to $n$, i.e., $\Sigma^{n} =\{ w \in \Sigma^{*} \mid |w| \leq n\} $. 
\begin{Definition}[Selection]
We define $(n,\ny,\udim)$-selection as a pair 
$\left(\alpha, \beta \right)$ such that 
\begin{enumerate}
	\item $\alpha \subseteq \Sigma^{n} \times \{1,2,\cdots, \ny \} $ and $\beta \subseteq \Sigma \times \Sigma^{n} \times \{1,2,\cdots,\udim \} $
	\item $\mathrm{card}(\alpha) = \mathrm{card}(\beta) = n $, where $\mathrm{card}$ denotes cardinality of the set.
\end{enumerate}
When $\ny$ and $\udim$ are clear from the context, we refer to $(n,\ny,\udim)$-selections as $n$-selections,
and when $n$ is also clear from the context, we use the term selection.
\end{Definition}
We will fix the following ordering of $\alpha$ and $\beta$.
\begin{equation}
\label{selection:enum}
\begin{split}
&\alpha = \{(u_{i},k_{i}) \}_{i=1}^{n}, ~ \beta = \{(\sigma_{j}, v_{j},l_{j})\}_{j=1}^{n},
\end{split}
\end{equation}
$u_{i} \in \Sigma^{n}$, $k_{i} \in \{1,2,\cdots,\ny \}$, $\sigma_{j} \in \Sigma$, $v_{j} \in \Sigma^{n}$, $l_{j} \in \{1,2,\cdots,\udim\}$
\begin{Example}\label{eg:selection}
	Consider $n\!=\!2$, number of outputs and inputs $\ny= \udim= \!\!\!=\!\!\!2$, and scheduling signal dimension $\pdim \!\!\!=\!\!\!2$, we have, $\Sigma^{n} = \{\epsilon,1,2,11,12,21,22\}$. Then, one of the \emph{$n$-selection} pair  $\left(\alpha, \beta \right)$  can be chosen as, for e.g.,
	$\alpha =\{\left(u_{1},k_1 \right), \left(u_2,k_2 \right)  \}= \{\left(\epsilon,1 \right), \left(11,2 \right)  \}$ and $\beta = \{\left(\sigma_{1},v_{1},l_1 \right), \left(\sigma_{2},v_2,l_2 \right)  \}=\{\left(1,21,1 \right), \left(2,22,2 \right)  \}$. 
\end{Example}


 

Let $M:\Sigma^{*} \rightarrow \mathbb{R}^{\ny \times \udim}$ be a map, values of which represent potential sub-Markov parameters \eqref{eqn:sub_markov} of an LPV-SSA.
Let us now define the Hankel matrix $\hankredu^{M} \in \mathbb{R}^{n \times n}$ as follows:
$i,j=1,\ldots,n$,
the $(i,j)$-th element of $\hankredu^{M}$ is of the form
\begin{equation}
\label{eqn:hankel_u}
\left[ \hankredu^{M} \right]_{i,j} \!=\! \left[M(\sigma_{j}v_{j} u_{i} ) \right]_{k_i,l_j}, 
\end{equation}
$\left[M(\sigma_{j}v_{j} u_{i} ) \right]_{k_i,l_j}$ denotes the entry of $M(\sigma_{j}v_{j} u_{i} )$ on the
$k_i$-th row and $l_j$-th column, and
$\left(u_{i},k_i\right) \in \alpha, \left(\sigma_{j},v_j,l_j\right) \in \beta$ are as in  the ordering of \eqref{selection:enum}. Intuitively, the rows of $\hankredu^M$ are indexed by word-index pairs $\left(u_i,k_i \right) \in \alpha$, where $u_i \in \Sigma^{n}$ and $k_i \in \{1,\ldots, \ny\}$ and similarly, the columns of  $\hankredu^M$ are indexed by word-index pairs $\left(\sigma_{j}v_j,l_j \right) \in \beta$, where $\sigma_{j} \in \Sigma$, $v_j \in \Sigma^{n}$ and $l_j \in \{1,\ldots, \udim\}$, and 
the element of $\hankredu^M$ with the row indexed $(u_i,k_i)$ and column index $(\sigma_{j},v_j,l_j)$ is the $(k_i,l_j)$-th entry
of $M(\sigma_{j}v_ju_i)$.  

In addition, we define the $\sigma$-shifted Hankel-matrix
$\mathcal{H}_{\sigma, \alpha, \beta}^M \in \mathbb{R}^{n \times n}$ as follows: its $i,j$-th entry is given by 
\begin{equation}\label{eqn:hankelu:shift}
\left[\mathcal{H}_{\sigma, \alpha, \beta}^M\right]_{i,j} = \left[ M(\sigma_{j}v_j \sigma u_i) \right]_{k_i,l_j}. 
\end{equation}
Moreover, let us define Hankel matrices   $\mathcal{H}_{\alpha,\sigma}^M \in \mathbb{R}^{n \times \udim}$ and $\mathcal{H}^M_{\beta} \in \mathbb{R}^{\ny \times n}$ as follows
\begin{align}
\left[\mathcal{H}^M_{\alpha, \sigma }\right]_{i,j} &= \left[ M(\sigma u_i)\right]_{k_i, j},  \ j=1,\ldots, \udim  \label{eqn:hankel_u_2a}\\
\left[\mathcal{H}^M_{\beta}\right]_{i,j} &= \left[ M(\sigma_j v_j) \right]_{i, l_j},  \ i=1,\ldots, \ny \label{eqn:hankel_u_2b}
\end{align}



Consider the model matrix computations   summarized in Algorithm \ref{algo:Nice_select_true_deter_basic}, using Hankel matrices and selections.

\begin{algorithm}[h!]
	\caption{Deterministic realization: Matrix computations  using Hankel matrices and $n$-selection}
	\label{algo:Nice_select_true_deter_basic}
	~~\textbf{Input}:  $(n,\ny,\udim)$-selection $\left(\alpha, \beta \right)$;  Hankel matrix $\hankredu^M$ \eqref{eqn:hankel_u}; shifted Hankel-matrix $\mathcal{H}^M_{\sigma, \alpha, \beta} $ \eqref{eqn:hankelu:shift}; Hankel matrices   $\mathcal{H}_{\alpha,\sigma}^M$ and $\mathcal{H}_{\beta}^M$ defined in \eqref{eqn:hankel_u_2a}-\eqref{eqn:hankel_u_2b} respectively, and  $M(\epsilon)$.
	\hrule\vspace*{.1cm}
	\begin{enumerate}[label=\arabic*., ref=\theenumi{}]
	\item  Compute the matrices 
               \begin{align*}
              	&  \hat{A}_{\sigma} = (\hankredu^M)^{-1}  \mathcal{H}_{\sigma, \alpha, \beta}^M, \\
              	&  \hat{B}_{\sigma} = (\hankredu^M)^{-1} \mathcal{H}^M_{\alpha,\sigma}, \ \hat{C} =  \mathcal{H}^M_{\beta}
               \end{align*}
		\item Compute $\hat{D}=M(\epsilon)$.
	\end{enumerate}
	\hrule \vspace*{.1cm}
	\textbf{Output}: Matrices $(\{\hat{A}_{\sigma}, \hat{B}_{\sigma} \}_{\sigSet}, \hat{C}, \hat{D} )$
\end{algorithm}
\begin{Lemma}[Adapted from \citep{CoxLPVSS}]
\label{basis_red:lemma}
Let the $(n,\ny,\udim)$-selection $(\alpha,\beta)$ be such that $\text{rank}(\hankredu^M) = n$, and assume that there exists a deterministic LPV-SSA representation $\mathscr{S}_{*}$ of dimension $n$
such that $M=M_{\mathscr{S}_{*}}$. Then the tuple $\hat{\mathscr{S}}=(\{\hat{A}_{\sigma}, \hat{B}_{\sigma} \}_{\sigSet}, \hat{C}, \hat{D})$, returned by Algorithm \ref{algo:Nice_select_true_deter_basic}, when applied to
the matrices $\hankredu^M$, $\mathcal{H}^M_{\sigma, \alpha, \beta}$, $\mathcal{H}_{\alpha,\sigma}^M$, $\mathcal{H}_{\beta}^M$ (\eqref{eqn:hankel_u}-\eqref{eqn:hankel_u_2b}) and $M(\epsilon)$, is a minimal dimensional deterministic LPV-SSA representation
such that $M_{\hat{\mathscr{S}}}=M$, i.e.
$M(\sigma w)=\hat{C}\hat{A}_w\hat{B}_{\sigma}$ for all $w \in \Sigma^{*}$. 
\end{Lemma}


\subsection{Correlation analysis: finding an LPV-SSA representation of $\yb^d$}
\label{real:alg1}
In this section, we describe an adaptation of the \emph{correlation
analysis} (CRA) method \citep{CoxIFAC,CoxLPVSS} for finding a stationary LPV-SSA representation of $\yb^d$ with noise process
$\ub$. 



Let us define the map  $\incov: \Sigma^{*} \rightarrow \mathbb{R}^{\ny \times \udim}$ as follows
\begin{equation}\label{eqn:input_covar}
\incov(w) =\left\{\begin{array}{lr} \frac{1}{\sqrt{p_w}} E[\yb(t)(\zwu(t))^{T} ] \varu^{-1} \ \ &  \forall w \in \Sigma^{+}  \\
                        E[\yb(t)\ub^T(t)]\varu^{-1} & w=\epsilon 
          \end{array}\right. 
\end{equation}
where we recall from Assumption \ref{asm:main}, $\varu \!=\! \mathrm{var}(\ub)$.

It turns out that if $\yb$ has a stationary LPV-SSA representation with input $\ub$, then $\incov$ is the sub-Markov function of a deterministic LPV-SSA representation.
\begin{Lemma}
\label{thm:cra}
Assume that $\yb$ has a realization by a stationary LPV-SSA representation with input $\ub$. 
Assume that $(\{A_{\sigma},B_{\sigma}\},C,D,\xb,\ub)$ is a stationary LPV-SSA representation (without inputs, Definition \ref{defn:LPV_SSA_wo_u}) of $\yb^d$. Then
$\incov$  in \eqref{eqn:input_covar} equals the sub-Markov function $M_{\mathscr{S}}$ \eqref{eqn:sub_markov} of the deterministic LPV-SSA representation
$\mathscr{S}=(\{A_{\sigma},B_{\sigma}\}_{\sigma \in \Sigma},C,D)$.
Conversely, if 
$\hat{\mathscr{S}}=(\{\hat{A}_{\sigma},\hat{B}_{\sigma}\}_{\sigma \in \Sigma},\hat{C},\hat{D})$ is a deterministic LPV-SSA representation 
such that its sub-Markov function $M_{\hat{\mathscr{S}}}$ equals $\incov$ and it is minimal dimensional among such deterministic LPV-SSA representations, then 
$(\{\hat{A}_{\sigma},\hat{B}_{\sigma}\},\hat{C},\hat{D},\hat{\xb},\ub)$ is a stationary LPV-SSA representation (without inputs) of $\yb^d$. 
\end{Lemma}
The proof of Lemma \ref{thm:cra} is presented in \cite[Appendix \ref{app:proof_lem4}]{MP_LPVS19_Arx}.

%
Hence,  we can adapt the \emph{basis reduced} Ho-Kalman realization algorithm as described in Algorithm \ref{algo:Nice_select_true_deter}.
\begin{algorithm}[h!]
	\caption{Realization of $\yb^d$: Computing an LPV-SSA representation of $\yb^d$  using covariances and $n$-selection.}
	\label{algo:Nice_select_true_deter}
	~~\textbf{Input}:  $(n,\ny,\udim)$-selection $\left(\alpha, \beta \right)$ of the form \eqref{selection:enum};  
            $\incov(\sigma_jv_ju_i)_{k_i,l_j}$, $\incov(\sigma u_i)_{k_i,l}$, $\incov(\sigma_j v_j)_{r,j}$,  $i,j=1,\ldots,\nx$, $l=1,\ldots,\udim$, $r=1,\ldots,\ny$, $\incov(\epsilon)$. 
	\vspace*{.1cm}\hrule\vspace*{.1cm}
	\begin{enumerate}[label=\arabic*., ref=\theenumi{}]
        \item
           Construct the matrices $\hankredu^{\incov}$, $\mathcal{H}^{\incov}_{\sigma, \alpha, \beta}$, $\mathcal{H}_{\alpha,\sigma}^{\incov}$ and $\mathcal{H}_{\beta}^{\incov}$, by replacing $M$ with $\incov$ in 
\eqref{eqn:hankel_u}--\eqref{eqn:hankel_u_2b}.
        \item Apply Algorithm \ref{algo:Nice_select_true_deter_basic}  to
               $\hankredu^{\incov}$, $\mathcal{H}^{\incov}_{\sigma, \alpha, \beta}$, $\mathcal{H}_{\alpha,\sigma}^{\incov}$, $\mathcal{H}_{\beta}^{\incov}$, $M(\epsilon)=\incov(\epsilon)$.
               Denote by $(\{\hat{A}_{\sigma}, \hat{B}_{\sigma} \}_{\sigSet}, \hat{C}, \hat{D})$
               the matrices returned by  Algorithm \ref{algo:Nice_select_true_deter_basic}. 
        \end{enumerate}
	\hrule \vspace*{.1cm}
	\textbf{Output}: Matrices $(\{\hat{A}_{\sigma}, \hat{B}_{\sigma} \}_{\sigSet}, \hat{C}, \hat{D})$
\end{algorithm}
It is clear from Lemma \ref{thm:cra} and  Lemma \ref{basis_red:lemma} that Algorithm \ref{algo:Nice_select_true_deter} is correct.
\begin{Corollary}
\label{thm:cra:col}
  If $\yb^d$ has a stationary LPV-SSA representation with no inputs, with noise process $\ub$, with dimension
$n$ and $\mathrm{rank}$ $ \hankredu^{\incov}=n$, then Algorithm \ref{algo:Nice_select_true_deter} returns matrices
$(\{\hat{A}_{\sigma}, \hat{B}_{\sigma} \}_{\sigSet}, \hat{C}, \hat{D})$ such that
$(\{\hat{A}_{\sigma}, \hat{B}_{\sigma} \}_{\sigSet}, \hat{C}, \hat{D}, \hat{\xb},\ub)$ is a stationary
LPV-SSA representation of $\yb^d$ without inputs, with noise process $\ub$.
\end{Corollary}


\subsection{Covariance realization algorithm}
\label{real:alg2}

In this section, we adapt the realization algorithm from \cite{MP_CDC19} to estimate the stochastic part \eqref{eqn:LPV_SSA_stoch} of a LPV-SSA representation.




Define the \emph{covariance sequence} $\Psi_{\ybs}: \Sigma^{*} \rightarrow \mathbb{R}^{\ny \times \ny}$, where $\Psi_{\ybs}(\epsilon) =I_{\ny}$, and for all $w \in \Sigma^{+}$,
\begin{equation}\label{eqn:covseq}
\covseq(w) = 
                     E[\ybs(t) (\zwys(t))^{T} ]  
\end{equation}
If $\yb^s$ has a stationary LPV-SSA representation, then $\covseq$ is a sub-Markov function of a
suitable deterministic LPV-SSA representation, \cite{PetreczkyBilinear,MP_CDC19}. 

Conversely, from a deterministic LPV-SSA representation, sub-Markov function of which equals $\covseq$ a 
stationary LPV-SSA representation can be computed. 
\begin{Lemma}
\label{lemma:stoch-real1}
If $\mathscr{S}=(\{\hat{A}_{\sigma},\hat{\Bs}_{\sigma}\}_{\sigma \in \Sigma},\hat{C},I_{\ny})$
is a minimal dimensional deterministic LPV-SSA representation such that $M_{\mathscr{S}}=\covseq$, then 
$(\{\hat{A}^s_{\sigma},\hat{K}_{\sigma}\}_{\sigma \in \Sigma},\hat{C},I_{\ny},\hat{\xb},\eb^s)$
is a stationary LPV-SSA representation of $\yb^s$ in forward innovation form, where
$\hat{A}^{s}_{\sigma}=\frac{1}{\sqrt{p_{\sigma}}} \hat{A}_{\sigma}$, $\hat{C}^{s}=\hat{C}_{\sigma}$, 
$\hat{K}_{\sigma}=\lim_{i \rightarrow \infty} \hat{K}_{\sigma}^{i}$, and
$\{\hat{K}_{\sigma}^i\}_{\sigma \in \Sigma, i \in \mathbb{N}}$ satisfies the following recursion
	\begin{equation}
  \label{statecov:iter}
  \begin{split}
	& 	\hat{P}^{i+1}_{\sigma} = \sum \limits_{\sigma_{1} \in \Sigma} \psig \left( \hat{A}^{s}_{\sigma_{1}} \hat{P}^{i}_{\sigma_{1} }  (\hat{A}^{s}_{\sigma_{1}})^{T}  + \hat{K}_{\sigma_{1}} \hat{Q}^{i}_{\sigma_{1} }  \hat{K}^{T}_{\sigma_{1}}  \right)\\
		& \hat{Q}^{i}_{\sigma} = \psig E[\zwsigs(t)(\zwsigs(t))^{T}] - \hat{C}^{s}  \hat{P}^{i}_{\sigma } (\hat{C}^{s})^{T}  \\ 
		& \hat{K}^{i}_{\sigma} = \left( \hat{\Bs}_{\sigma} \sqrt{\psig} - \hat{A}^{s}_{\sigma} \hat{P}^{i}_{\sigma}  (\hat{C}^{s})^{T} \right) \left(  \hat{Q}^{i}_{\sigma} \right)^{-1}
		\end{split}
\end{equation}
 with $\hat{P}^{0}_{\sigma} =0$. 
Moreover, 
$E[\eb^s(t)(\eb^s(t))^T\p^2_{\sigma}(t)]=\hat{Q}_{\sigma}=\lim_{i \rightarrow \infty} \hat{Q}_{\sigma}^i$,
$E[\hat{\xb}(t)\hat{\xb}^T(t)\p^2_{\sigma}]=\hat{P}_{\sigma}=\lim_{i \rightarrow \infty} \hat{P}_{\sigma}^i
$ for all $\sigma \in \Sigma$. 
\end{Lemma}
The proof of Lemma \ref{lemma:stoch-real1} can be found in \cite{PetreczkyBilinear,MP_CDC19}, \cite[Appendix \ref{app:proof_lem5}]{MP_LPVS19_Arx}.
From Lemma \ref{lemma:stoch-real1}, it follows that we can use the basis reduced Kalman-Ho realization algorithm
Algorithm \ref{algo:Nice_select_true_deter}, as described in Algorithm \ref{algo:Nice_select_true}, in order to compute LPV-SSA representation of $\yb^s$ .

\begin{algorithm}[h!]
	\caption{Realization of $\yb^s$: Computing an LPV-SSA representation of $\yb^s$  using covariances and $n$-selection.}
	\label{algo:Nice_select_true}
	~~\textbf{Input}:  $(n,\ny,\ny)$-selection $\left(\alphas, \betas \right)$ of the form  of the form \eqref{selection:enum}; 
        $\covseq(\sigma_jv_ju_i)_{k_i,l_j}$; $\covseq(\sigma u_i)_{k_i,l}$; $\covseq(\sigma_j v_j)_{r,j}$, $i,j=1,\ldots,n$; 
        $l, r=1,\ldots,\ny$, $\{E[\zwsigs(t)(\zwsigs(t))^{T}]\}_{\sigma \in \Sigma}$; number of maximal iterations $\mathcal{I} > 0$.
	\vspace*{.1cm}\hrule\vspace*{.1cm}
	\begin{enumerate}[label=\arabic*., ref=\theenumi{}]
        \item
           Construct the matrices $\mathcal{H}^{\covseq}_{\alphas, \betas}$, $\mathcal{H}^{\covseq}_{\sigma, \alphas, \betas}$, $\mathcal{H}_{\alphas,\sigma}^{\covseq}$, $\mathcal{H}_{\betas}^{\covseq}$ by replacing $M$ with $\covseq$ in \eqref{eqn:hankel_u}--\eqref{eqn:hankel_u_2b}.
        \item Apply Algorithm \ref{algo:Nice_select_true_deter_basic}  to
               $\mathcal{H}^{\covseq}_{\alphas, \betas}$, $\mathcal{H}^{\covseq}_{\sigma, \alphas, \betas}$, $\mathcal{H}_{\alphas,\sigma}^{\covseq}$, $\mathcal{H}_{\betas}^{\covseq}$, $M(\epsilon)=I_{\ny}$.
               Denote by $(\{\hat{A}_{\sigma}, \hat{\Bs}_{\sigma} \}_{\sigSet}, \hat{C}, I_{\ny})$
               the matrices returned by  Algorithm \ref{algo:Nice_select_true_deter_basic}. 
        \item Define $\hat{A}_{\sigma}^s=\frac{1}{\sqrt{p_\sigma}} \hat{A}_{\sigma}$, $\sigma \in \Sigma$,
              $\hat{C}^s=\hat{C}$. 
        \item Compute $\{\hat{K}_{\sigma}^{i},\hat{Q}^i_{\sigma},\hat{P}_{\sigma}^i\}_{i=1}^{\mathcal{I}}$ using the recursion \eqref{statecov:iter}. 
        \end{enumerate}

	\hrule \vspace*{.1cm}
	\textbf{Output}: Matrices $(\{\hat{A}^{s}_{\sigma}, \hat{\Bs}_{\sigma}, \hat{K}^{\mathcal{I}}_{\sigma}, \hat{Q}^{\mathcal{I}}_{\sigma}, \hat{P}^{\mathcal{I}}_{\sigma} \}_{\sigSet}, \hat{C}^{s})$
\end{algorithm}
It is clear from Lemma \ref{lemma:stoch-real1} and  Lemma \ref{basis_red:lemma} that Algorithm \ref{algo:Nice_select_true} is correct.
\begin{Corollary}
\label{thm:cra:col1}
  If $\yb^s$ has a stationary LPV-SSA representation with no inputs, with dimension
$n$ and $\mathrm{rank} \ \hankred^{\incov}\!\!=\!\!n$, then Algorithm \ref{algo:Nice_select_true} returns matrices
$(\{\hat{A}^{s}_{\sigma}, \hat{\Bs}_{\sigma}, \hat{K}^{\mathcal{I}}_{\sigma}, \hat{Q}^{\mathcal{I}}_{\sigma}, \hat{P}^{\mathcal{I}}_{\sigma} \}_{\sigSet}, \hat{C}^{s})$ such that with $\hat{K}_{\sigma}=\lim_{\mathcal{I} \rightarrow \infty} \hat{K}_{\sigma}^{\mathcal{I}}$, $\hat{Q}_{\sigma}=\lim_{\mathcal{I} \rightarrow \infty} \hat{Q}_{\sigma}^{\mathcal{I}}$,
$\hat{P}_{\sigma}=\lim_{\mathcal{I} \rightarrow \infty} \hat{P}_{\sigma}^{\mathcal{I}}$;
 tuple $(\{\hat{A}_{\sigma}^s, \hat{K}_{\sigma} \}_{\sigSet}, \hat{C}^s, I_{\ny}, \hat{\xb},\eb^s)$ is a stationary
LPV-SSA representation of $\yb^s$ without inputs, and $\hat{Q}_{\sigma}\!\!=\!\!E[\eb^s(t)(\eb^s(t))^T\p_{\sigma}^2(t)]$,
$\hat{P}_{\sigma}\!\!=\!\!E[\hat{\xb}(t)\hat{\xb}^T(t)\p_{\sigma}^2(t)]$, $\sigma \in \Sigma$. 
\end{Corollary}

\section{Identification algorithm}
\label{sect:ident}

In this section, we formulate an identification algorithm  based on stochastic realization Algorithms \ref{algo:Nice_select_true_deter}--\ref{algo:Nice_select_true} and selections,  for  $N$-length observation sequence of outputs, inputs and scheduling signals, as detailed in Algorithm~\ref{algo:Nice_select}. 
Intuitively, the main idea behind Algorithm~\ref{algo:Nice_select} is to estimate the covariances $\incov$, $\covseq$ and  $E[\z^{\yb}_{\sigma}(t)(\z^{\yb}_{\sigma}(t))^{T} ]$
from the observed data and then apply  Algorithms \ref{algo:Nice_select_true_deter}--\ref{algo:Nice_select_true}  to the thus estimated covariances. 
More specifically, the following assumptions are made: 
\begin{Assumption}
	\label{assum1}
	
	\textbf{(1)}
	The $\nx$-selection pair $\left(\alpha, \beta \right)$ and  $\left(\alphas, \betas \right)$    are such that 
	$ \mathrm{rank} \ \hankredu^{\incov} = \nx$,  
	$ \mathrm{rank} \ \mathcal{H}_{\alphas,\betas}^{\covseq} = \nx$, where $\nx$ is the state-space dimension of a minimal LPV-SSA realization of $\yb$.
	
	\textbf{(2)}
	The process $(\yb, \ub,  \{\p_{w} \}_{\wordSet{+}})$ is ergodic and there exist sample paths
	$y: \mathbb{Z} \rightarrow \mathbb{R}^{\ny}$, $u: \mathbb{Z} \rightarrow \mathbb{R}^{\udim}$ and $\mu: \mathbb{Z} \rightarrow \mathbb{R}^{\pdim}$ of the processes $\yb$, $\ub$ and $\p$ respectively such that
	$\{y(t),u(t), \{{\mu}_{\sigma}(t)\}_{\sigSet} \}_{t=1}^{N}$ is observed and the following holds:
        for all $w \in \Sigma^{*}, \sigma \in \Sigma$, 
	\begin{equation*}
        \begin{split}
 & {\incov^N(w)}= \frac{1}{\sqrt{p_w}} 
\left( \frac{1}{N} \sum_{t=|w|}^{N} y(t)(z_w^{u}(t))^{T}  \right)\varu^{-1},  ~ w \in  \Sigma^{*} \\
& \Lambda^{\yb,N}_{\sigma w}=\frac{1}{N} \sum_{t=|w|}^{N} y(t)(z_{\sigma w}^{y}(t))^{T}, ~  T^{\yb,N}_{\sigma,\sigma}=\frac{1}{N} \sum_{t=1}^{N} z^{y}_{\sigma}(t)(z_{\sigma}^{y}(t))^{T} 
\end{split}
\end{equation*}
Then for all $w \in \Sigma^{*}$, $\sigma \in \Sigma$,
\begin{align*}
& \incov(w) =  \lim\limits_{N \rightarrow \infty} \incov^N(w),   ~ E[\z^{\yb}_{\sigma}(t)(\z^{\yb}_{\sigma}(t))^{T}]=\lim\limits_{N \rightarrow \infty} T^{\yb,N}_{\sigma,\sigma} \\
& E[\yb(t)(\z^{\yb}_{\sigma w}(t))^T] =  \lim\limits_{N \rightarrow \infty} \Lambda^{\yb,N}_{\sigma w}, ~ 
\end{align*}
	where, 
	for all $w=\sigma_1\sigma_2\cdots \sigma_r \in \Sigma^{+}$, $r > 0$, we have,
	\begin{equation*}
	\begin{split} 
	& \mu_{w}(t) = \mu_{\sigma_{1}}(t-k+1)\mu_{\sigma_{2}}(t-k+2)\cdots \mu_{\sigma_{r}}(t) \\
	& z^{u}_w(t) = u(t-|w|) \mu_{w}(t-1)\frac{1}{\sqrt{p_{w}}}, ~  z^u_{\epsilon}(t)=u(t) \\
        &  z^{y}_w(t) = y(t-|w|) \mu_{w}(t-1)\frac{1}{\sqrt{p_{w}}}
	\end{split}
	\end{equation*}
	
\end{Assumption}



\begin{algorithm}[h]
	\caption{Identification of stochastic LPV-SSA from observed data.}
	\label{algo:Nice_select}
	\textbf{Input}: Observations sequence  $\{{y}(t),u(t), \{{\mu}_{\sigma}(t)\}_{\sigSet} \}_{t=1}^N$, and $\nx$-selection $\left(\alpha, \beta \right)$ and $\left(\alphas, \betas \right)$; $\{\psig\}_{\sigSet}$, $\varu$, 
	maximum number of iterations $\mathcal{I} > 0$.
	\vspace*{.1cm}\hrule\vspace*{.1cm}
	\begin{enumerate}[label=\arabic*., ref=\theenumi{}]
		\item Compute empirical covariances $\incov^N(w)$,
		for every $w \in \Sigma^{+}$, such that $w=ivu$ or  $w=i v\sigma u$ or  $w=iv$ or $w=\sigma u$ for some words $v,u \in \Sigma^{*}$, $\sigma \in \Sigma$,
		$(u,k) \in \alpha$, $(i, v,l) \in \beta$ for some $k=1,\ldots,\ny$, $l=1,\ldots,\udim$, 
		$ \forall w \in \Sigma^{+}$. 
		
		\item Run Algorithm \ref{algo:Nice_select_true_deter} with empirical covariances $\incov^N(w)$, instead of 
                      the covariances $\incov$.
    Denote the result returned by  Algorithm~\ref{algo:Nice_select_true_deter}
		by $\mathscr{S}=(\{\tilde{A}_{\sigma}^{d}, \tilde{B}_{\sigma}^{d} \}_{\sigSet}, \tilde{C}^{d},\tilde{D}^{d} )$.
		
		
%
		 
		\item Compute approximate covariances:
\begin{equation}
\label{covseq:lemma:eq1}
 \begin{split}
  & \covseq^N(\sigma w)=\Lambda_{\sigma w}^{\yb,N}-\Lambda_{\mathscr{S}}(\sigma w) \\
  & \Lambda_{\mathscr{S}}(\sigma w)=\frac{1}{\sqrt{p_{\sigma w}}} \tilde{C}^d \tilde{A}_{w}^d(\tilde{A}^d_{\sigma} \tilde{\Psig} (\tilde{C}^d)^{T} + \tilde{B}^d_{\sigma} \varu ) \\
  & \Tcov^N=T^{\yb,N}_{\sigma,\sigma}-T_{\sigma,\sigma,\mathscr{S}} \\
  & 
T_{\sigma,\sigma,\mathscr{S}} = \frac{1}{\psig} (\tilde{C}^d\tilde{\Psig} (\tilde{C}^d)^{T} + \varu) 
 \end{split}
 \end{equation}
 for all $\sigSet$ and for every $w \in \Sigma^{+}$, such that $w=ivu$ or $w=iv$ or $w=iu$ or $w=i v\sigma u$ for some words $v,u \in \Sigma^{*}$, $i,\sigma \in \Sigma$,
		$(u,k) \in \alphas $, $(v,l) \in \betas $ for some $k,l=1,\ldots,\ny$, 
		for all $w \in \Sigma^{+}$.
  Here, $\tilde{\Psig}$ is the unique solution to the following Sylvester equation
  \begin{equation}
 \label{covseq:lemma:eq2}
 \tilde{\Psig} \!\!=\!\! \psig \sum \limits_{\sigma_{1} \in \Sigma} \left( \tilde{A}^d_{\sigma_{1}} \tilde{P}_{\sigma_{1} }  (\tilde{A}^d_{\sigma_{1}})^{T}  + \tilde{B}^{d}_{\sigma_{1}} \varu  (\tilde{B}^{d}_{\sigma_{1}})^T  \right).
 \end{equation}
		 
		
		
		\item Run Algorithm~\ref{algo:Nice_select_true}, with 
                      the empirical covariances $\covseq^N$ instead of $\covseq$, and $\{\Tcov^{N}\}_{\sigma \in \Sigma}$ instead of
                     $\{E[\zwsigs(t)(\zwsigs(t))]^{T}\}_{\sigma \in \Sigma}$.
Denote the result returned by  Algorithm~\ref{algo:Nice_select_true}
		by $(\{\tilde{A}^{s}_{\sigma}, \tilde{K}^{\mathcal{I},s}_{\sigma}, \tilde{Q}^{\mathcal{I}}_{\sigma}, \tilde{P}^{\mathcal{I}}_{\sigma} \}_{\sigSet}, \tilde{C}^{s} )$.
		
\item The estimated model matrices of LPV-SSA \eqref{eqn:LPV_SSA} are given by  $\tilde{A}_{\sigma}^N = \mathrm{diag}(\tilde{A}^{d}_{\sigma},\tilde{A}^{s}_{\sigma})$, $\tilde{B}^N_{\sigma} = \left[ (\tilde{B}^{d}_{\sigma})^{T} \ 0^{T} \right]^{T}$, $\tilde{K}^{N,\mathcal{I}}_{\sigma} = \left[ 0^{T} \  (\tilde{K}^{\mathcal{I},s}_{\sigma})^{T}  \right]^{T}$, $\forall \sigSet$, $\tilde{C}^N = \left[ \tilde{C}^{d}  \ \ \tilde{C}^{s}  \right]$, $\tilde{D}^N =  \tilde{D}^{d}$,  $\tilde{Q}^{N,\mathcal{I}}_{\sigma}=\tilde{Q}^{\mathcal{I}}_{\sigma}$.


		
	\end{enumerate}
	\hrule\vspace*{.1cm}
	\textbf{Output}: 
	Estimates $\{\tilde{A}_{\sigma}^N, \tilde{B}^N_{\sigma}, \tilde{K}^{N,\mathcal{I}}_{\sigma}, \tilde{Q}^{N,\mathcal{I}}_{\sigma} \}_{\sigSet}, \tilde{C}^N, \tilde{D}^N$
\end{algorithm}
\begin{Lemma}[Consistency]
\label{lm:consistency}
   With the Assumption \ref{assum1}
   the result of Algorithm \ref{algo:Nice_select} satisfies the following:  
    \begin{align*}
     & \tilde{K}_{\sigma}= \lim_{\mathcal{I} \rightarrow \infty} \lim_{N \rightarrow \infty}\tilde{K}^{N,\mathcal{I}}_{\sigma}, ~
      \tilde{A}_{\sigma}=\lim_{N \rightarrow \infty} \tilde{A}_{\sigma}^{N}, ~
\\
     & \tilde{B}_{\sigma}=\lim_{N \rightarrow \infty} \tilde{B}_{\sigma}^{N}, ~  \tilde{C}=\lim_{N \rightarrow \infty} \tilde{C}^{N}, ~ 
     \tilde{D}=\lim_{N \rightarrow \infty} \tilde{D}^{N}
   \end{align*}
 and $(\{\tilde{A}_{\sigma},\tilde{B}_{\sigma},\tilde{K}_{\sigma},\}_{\sigma=1}^{\pdim},\tilde{C},\tilde{D},\hat{\xb},\eb^s)$ is a stationary LPV-SSA representation of $(\yb,\ub,\p)$,
and $E[\eb^s(t)(\eb^s(t))^T\p_{\sigma}^2(t)]=\lim_{\mathcal{I} \rightarrow \infty} \lim_{N \rightarrow \infty} \tilde{Q}^{N,\mathcal{I}}_{\sigma}$,
$\sigma \in \Sigma$.
\end{Lemma}
The proof sketch of Lemma \ref{lm:consistency} is presented in \cite[Theorem 3]{MP_CDC19}, \cite{MP_LPVS19_Arx}.
\begin{Remark}[Intuition behind \eqref{covseq:lemma:eq1}]
 It can be shown that 
 $\covseq(\sigma w)=E[\yb(t)(\z_{\sigma w}^{\yb}(t))^T]-E[\yb^d(t)(\z_{\sigma w}^{\yb^d}(t))^T]$ and
 $E[\zwsigs(t)(\zwsigs(t))^{T}]=E[\z^{\yb}_{\sigma}(t)(\z^{\yb}_{\sigma}(t))^T]-E[\z^{\yb_d}_{\sigma}(t)(\z^{\yb_d}_{\sigma}(t))^T]$, see \cite{MP_LPVS19_Arx}.
 Moreover, if $(\{\tilde{A}_{\sigma}^{d}, \tilde{B}_{\sigma}^{d} \}_{\sigSet}, \tilde{C}^{d},\tilde{D}^{d},\hat{\xb}^d,\ub)$ is a stationary LPV-SSA representation of $\yb^d$ without inputs, then  from \cite{PetreczkyBilinear} it  follows that $\Lambda_{\mathscr{S}}(\sigma w)=E[\yb^d(t)(\z_{\sigma w}^{\yb^d}(t))^T]$ and
$T_{\sigma,\sigma,\mathscr{S}}=E[\z^{\yb^d}_{\sigma}(t)(\z^{\yb^d}_{\sigma}(t))^T]$. 
Intuitively, since $\incov^N(w)$ converges to $\incov(w)$ as $N \rightarrow \infty$, $(\{\tilde{A}_{\sigma}^{d}, \tilde{B}_{\sigma}^{d} \}_{\sigSet}, \tilde{C}^{d},\tilde{D}^{d},\hat{\xb}^d,\ub)$ becomes a LPV-SSA representation of $\yb^d$ as $N \rightarrow \infty$, and hence the right-hand side of the first and third equation of \eqref{covseq:lemma:eq1} converges
to $E[\yb(t)(\z_{\sigma w}^{\yb}(t))^T]-E[\yb^d(t)(\z_{\sigma w}^{\yb^d}(t))^T]$ and
$E[\z^{\yb}_{\sigma}(t)(\z^{\yb}_{\sigma}(t))^T]-E[\z^{\yb_d}_{\sigma}(t)(\z^{\yb_d}_{\sigma}(t))^T]$ respectively.
\end{Remark}
\begin{Remark}[Alternative way of computing $\covseq^N$]
\label{rem:alternative}
  An alternative way of estimating the covariances $\covseq$ and $E[\zwsigs(t)(\zwsigs(t))^{T}]\}_{\sigma \in \Sigma}$ is to use
  the matrices $\mathscr{S}=(\{\tilde{A}_{\sigma}^{d}, \tilde{B}_{\sigma}^{d} \}_{\sigSet}, \tilde{C}^{d},\tilde{D}^{d} )$ to approximate
  the sample paths $y^d$, $y^s$ of $\yb^d$ and $\ybs$ by 
  $\hat{y}^d(t)=\tilde{D}^du(t)+\sum_{v \in \Sigma^{*}, \sigma \in \Sigma,|v|  < t-1} \tilde{C}^d\tilde{A}_{v}^d\tilde{B}_{\sigma}^{d} z_{\sigma v}^{u}(t)$, and
  $\hat{y}^s(t)=y(t)-\hat{y}^d(t)$ and define 
  \begin{equation}
   \label{rem:alternative:eq1}
   \begin{split} 
  & \covseq^{N}(w) = \frac{1}{N} \sum_{t=|w|}^{N} \hat{y}^{s}(t) {z}^{\hat{y}^{s}}_{w}(t), ~ w \in \Sigma^{+} \\
  & \Tcov^{N} = \frac{1}{N} \sum_{t=|w|}^{N} {z}^{\hat{y}^{s}}_{\sigma}(t)({z}^{\hat{y}^{s}}_{\sigma}(t))^{T}, ~\sigma \in \Sigma
  \end{split}
  \end{equation}
   where ${z}^{\hat{y}^{s}}_{v}(t)=\hat{y}_s(t-|v|)\mu_v(t-1)\frac{1}{\sqrt{p_v}}$ for all $v \in \Sigma^{+}$. 
We can then view
 $\covseq^N(w)$ as an approximation of $\covseq(w)$, and  $\Tcov^{N}$ is an approximation of $E[\zwsigs(t)(\zwsigs(t))^{T}]\}_{\sigma \in \Sigma}$.
  We could modify Algorithm \ref{algo:Nice_select} by replacing \eqref{covseq:lemma:eq1} with \eqref{rem:alternative:eq1}. We conjecture that Lemma \ref{lm:consistency}
  will remain true for the modified algorithm. 
\end{Remark}

\section{Numerical example}\label{sec:example}
In this section, we present a numerical example to test the effectiveness of our algorithm. All computations are carried out on an i5 1.8-GHz Intel core processor with 8 GB of RAM running MATLAB R2018a.

The quality of the match between estimated and true outputs is quantified on a noise-free validation data of length $N_\mathrm{val}$ via  \emph{Best Fit Rate} (BFR) and \emph{Variance Accounted For} (VAF) criterion defined for each output channel $y_{i}$, $i\!=\!1,\ldots,\ny$, as
\begin{align*}
\mathrm{BFR}_{y_{i}} &=\max\left\{1-\sqrt{\frac{\sum_{t=1}^{N_\mathrm{val}}\left(y_{i}(t)-\hat{y}_{i}(t)\right)^2}{\sum_{t=1}^{N_\mathrm{val}}\left(y_{i}(t)-\bar{y_{i}}\right)^2}},0\right\} \times 100 \%,   \\
\mathrm{VAF}_{y_{i}} &=  \max\left\{1- \frac{\mathrm{var}(y_{i}- \hat{y}_{i})}{\mathrm{var}(y_{i})},0\right\} \times 100 \%, 
\end{align*}
where $\hat{y}_{i}$ denotes the simulated one-step ahead model output  and $\bar{y}_{i}$ denotes the sample mean of the output over the validation set. 

The LPV-SSA representation in form \eqref{eqn:LPV_SSA} is used for data generation with following matrices:
\[\begin{split}
& A_{1} = \begin{bmatrix}
0.4&0.4  &0 \\ 
0&0  &0 \\ 
0&0  &0 
\end{bmatrix}, 
A_{2} = \begin{bmatrix}
0&0  &0 \\ 
0&0.4  &0.4 \\ 
0&0.4  &0.4 
\end{bmatrix}, \\
&B_{1} = \left[1 \ 1 \ 1 \right]^{T}, 
B_{2} = \left[1 \ 0 \ 1 \right]^{T}, \\
&K_{1} = 
\left[-0.036 \ 0 \ 1  \right]^{T},
K_{2} = \left[0 \ \ 0.015 \ 1.17  \right]^{T},\\
&C = \begin{bmatrix}
1 & 0 & 0
\end{bmatrix},
D =1,
\end{split}
\]
which corresponds to state-dimension $\nx\!\!=\!\!3$, output dimension $\ny\!\!=\!\!1$, and scheduling signal dimension $\pdim\!\!=\!\!2$ with $\Sigma \!=\! \{1,2\}$. 
Note that, the system corresponding to first local model $\tilde{A}_{1} = A_{1}- K_{1}C$ is \emph{not observable}, i.e.,  $ \mathrm{rank}([C^{T} (C\tilde{A}_{1})^{T} \ldots (C\tilde{A}^{l-1}_{1})^{T}]^{T}) =2< \nx$, which is a particular assumption required in subspace based approaches \citep{vansubspace}. 

Training and noise free validation output sequences of length $N\!\!=\!\!100000$ and $N_\mathrm{val}\!\!=\!\!100000$, respectively, are generated using a white-noise input process $\ub$ with uniform distribution $\mathcal{U}(-1.5,1.5)$ and an independent scheduling signal process $\p = [\p_{1} \  \ \p_{2}] $ such that $\p_{1}(t) =1$ and $\p_{2}(t) $ is a white-noise process with uniform distribution $\mathcal{U}(-1.5,1.5)$. This corresponds to the parameter values $\{\psig \}_{\sigma \in \{1,2\}}$  to be $p_{1} \!=\! E[\mu^{2}_{1}(t)] \!\!= \!\!1$ and $p_{2} \!= \!E[\mu^{2}_{2}(t)]\!= \!0.75$. The standard deviation of the white Gaussian noise $\mathbf{e}$ corrupting the training output is $1$, i.e., $\eb \sim \mathcal{N}(0,1)$. This corresponds to the \emph{Signal-to-Noise Ratio} $\mathrm{SNR}=10\log{\frac{\sum_{t=1}^N\left(y(t)-e(t)\right)^2}{\sum_{t=1}^Ne^{2}(t)}}= 4.7 \ \mathrm{dB}.$ 

We run the version of Algorithm \ref{algo:Nice_select} explained in Remark \ref{rem:alternative}, with $\mathcal{I}=50$ iterations and with  the following $n$-selection pairs $(\alpha, \beta)$ and $(\alphas, \betas)$, with $n=3$, 
\begin{align*}
\alpha &= \{ (\epsilon,1), (1,1) , (21,1)\}, \ \beta = \{ (2,\epsilon,1), (1,2,1),(2,21,1)\}, \\
\alphas &= \{ (\epsilon,1), (1,1) , (21,1)\}, \ \betas = \{ (1,\epsilon,1), (1,2,1),(1,21,1)\}, 
\end{align*}
which are used to choose corresponding entries of the Hankel matrices. 
The mean time taken to run the algorithm is $1.55$ sec.


\begin{table}
	\caption{BFR and VAF on a noise-free validation data Algorithm \ref{algo:Nice_select}}\label{Tab:BFR_eg1}
	\begin{center} \vspace{-0.0cm}
		\begin{tabular}{|c||c|}  
			\hline
			BFR & 93.56 \% \\
			\hline
			VAF & 99.58 \% \\
			\hline
		\end{tabular} 
	\end{center}
\end{table}

\begin{table}
	\caption{True vs estimated sub-Markov parameters}\label{Tab:Markov}
	\begin{center} \vspace{-0.0cm}
		\begin{tabular}{|c||c|c|}  
				\hline
			Markov parameters	& True value & Estimated value\\
			\hline
		$CA_{1}B_{1}$	& 0.80 & 0.7957 \\
		$CA_{1}B_{2}$ & 0.40 & 0.3914 \\
		$CA^{2}_{1}B_{1}$ & 0.32 & 0.3147 \\
		$CA_{1}A_{2}B_{2}$& 0.16 & 0.1549 \\
		$CA^{3}_{1}B_{1}$ & 0.12 & 0.1093 \\
		\hline
		\end{tabular} 
	\end{center}
\end{table}

The validation result using one-step ahead predicted outputs $\hat{y}$  are reported in Table~\ref{Tab:BFR_eg1}, and true vs estimated sub-Markov parameters are reported in Table~\ref{Tab:Markov}. The results  show a good match between estimated model output w.r.t. true system output. 

\section{Conclusion}

In this paper, we formulated a realization algorithm and  an efficient identification algorithm  for stochastic LPV-SSA representations with inputs, by combining correlation analysis method with a stochastic realization based identification algorithm. The proposed algorithm  provides a computationally efficient alternative to the parametric subspace approaches avoiding the curse of dimensionality. 

\bibliography{StochLPVbib}
                                                   
\appendix
\section{Proofs}

\subsection{Proof of Lemma \ref{decomp:lemma}} \label{app:proof_lem1}
Recall from Notation \ref{hilbert:notation} the definition of the Hilbert-space $\mathcal{H}_1$ of zero mean square integrable random variables. 
Let us denote by $\mathcal{H}_{t,+}^{\ub}$, the closed subspace of $\mathcal{H}_1$ generated by the components of  $\{ \z^{\ub}_{w}(t) \}_{w \in \Sigma^{+}} \cup \{\ub(t)\}$.
 \begin{Lemma}
	\label{decomp:lemma:pf1}
	With the assumptions and notations  of Lemma \ref{decomp:lemma}, 
	$\vb(t)\p_{\sigma}(t)$ for all $\sigSet$, is orthogonal to $\mathcal{H}_{t,+}^{\ub}$. 
\end{Lemma}
\begin{pf}[Proof of Lemma \ref{decomp:lemma:pf1}]
	Since $\mathbf{r}(t):=\begin{bmatrix} \vb^T(t) & \ub^T(t) \end{bmatrix}^T$ is a white noise process w.r.t. $\p$, it follows that
	$E[\mathbf{r}(t+1)(\z^{\mathbf{r}}_{w}(t+1))^T]=0$ for all $w \in \Sigma^{+}$. 
	 In particular, as $\z^{\mathbf{r}}_{\sigma}(t+1)=\frac{1}{\sqrt{p}_{\sigma}} \mathbf{r}(t)\p_{\sigma}(t)$, $E[\mathbf{r}(t+1)(\mathbf{r}(t)\p_{\sigma}(t))^T]=0$ and as $E[\vb(t)(\ub(t+1))^T \p_{\sigma}(t)]$ is the transpose of the lower left block of $E[\mathbf{r}(t+1)(\mathbf{r}(t)\p_{\sigma}(t))^T]$, it follows that $E[\vb(t)(\ub(t+1))^T \p_{\sigma}(t)]=0$. 
	 
	 Since $\mathbf{r}$ is ZMWSSI, it follows that for all $w \in \Sigma^{+}$, $\sigma_1,\sigma \in \Sigma$
	 \begin{equation}
	 \label{decomp:lemma:pf1:eq1} 
	 \begin{split}
	 & E[\mathbf{r}(t)\p_{\sigma}(t)(\z_{w\sigma_1}^{\mathbf{r}}(t+1))^T]\\
	 &=\sqrt{p_{\sigma}}E[\z^{\mathbf{r}}_{\sigma}(t+1)(\z_{w\sigma_1}^{\mathbf{r}}(t+1))^T] \\
	 &= \left\{\begin{array}{cl}
	 0, & \mathrm{if} \ \ \sigma \ne \sigma_1 \\
	 \sqrt{p_{\sigma}}E[\mathbf{r}(t)(\z_{w}^{\mathbf{r}}(t))^T], & \mathrm{if} \ \ \sigma = \sigma_{1} 
	 \end{array}\right. 
	 \end{split}
	 \end{equation}
	 
	 Since $\mathbf{r}$ is a white noise process w.r.t. $\p$, it follows that $E[\mathbf{r}(t)\z_w^{\mathbf{r}}(t)]=0$ for all $w \in \Sigma^{+}$.
	 
	 Since $E[\vb(t)\p_{\sigma}(t)(\z^{\ub}_{w\sigma_1}(t+1))^T]$ is the upper right block of $E[\mathbf{r}(t)\p_{\sigma}(t)(\z^{\mathbf{r}}_{w\sigma_1}(t+1))^T]$,
	 it follows that $E[\vb(t)\p_{\sigma}(t)(\z^{\ub}_{w\sigma_1}(t+1))^T]=0$ for all $w \in \Sigma^{+}$, $\sigma \in \Sigma$. 
	 
	 That is, we have shown that  $E[\vb(t) \p_{\sigma}(t) (\ub(t+1))^T]=0$, $E[\vb(t)\p_{\sigma}(t)(\z^{\ub}_{w\sigma_1}(t+1))^T]=0$ for all $w \in \Sigma^{+}$, $\sigma, \sigma_1 \in \Sigma$. It is left to show that $E[\vb(t)\p_{\sigma}(t)(\z^{\ub}_{\sigma_1}(t+1))^T]=0$. Note that $E[\vb(t)\p_{\sigma}(t)(\z^{\ub}_{\sigma_1}(t+1))^T]$ is the 
	 upper right block of $p_{\sigma} E[\z^{\mathbf{r}}_{\sigma}(t+1)(\z_{w\sigma_1}^{\mathbf{r}}(t+1))^T]$, and the latter equals zero if $\sigma_1 \ne \sigma$.
	 Hence, for  $\sigma \ne \sigma_1$, $E[\vb(t)\p_{\sigma}(t)(\z^{\ub}_{\sigma_1}(t+1))^T]=0$. If $\sigma=\sigma_1$, then $E[\vb(t)\p_{\sigma}(t)(\z^{\ub}_{\sigma}(t+1))^T]=\frac{1}{\sqrt{p_{\sigma}}} E[\vb(t)\ub^T(t)\p_{\sigma}^2(t)]$, and 
	 from Definition \ref{def:Stationary} it follows that $E[\vb(t)\ub^T(t)\p_{\sigma}^2(t)]=0$, $\sigma \in \Sigma$. That is, $E[\vb(t)\p_{\sigma}(t)(\z^{\ub}_{\sigma}(t+1))^T]=0$. 
	 
	 To summarize, we have shown that $E[\vb(t)(\ub(t+1))^T\p_{\sigma}(t)]=0$, $E[\vb(t)\p_{\sigma}(t)(\z^{\ub}_{w\sigma_1}(t+1))^T]=0$ for all $w \in \Sigma^{*}$, $\sigma, \sigma_1 \in \Sigma$.
	 Since $\ub(t+1)$, $\z^{\ub}_{w\sigma_1}(t+1)$, $w \in \Sigma^{*}$, $\sigma_1 \in \Sigma$ generate $\mathcal{H}_{t+1,+}^{\ub}$, the statement of the lemma follows.
	 \hfill $ \blacksquare$
\end{pf}

Let us denote by $\mathcal{H}_{t}^{\ub}$, the closed subspace generated by the components of  $\{ \z^{\ub}_{w}(t) \}_{w \in \Sigma^{+}}$.
It is clear that $\mathcal{H}_{t}^{\ub} \subseteq \mathcal{H}_{t,+}^{\ub}$.

\begin{Lemma}
	\label{decomp:lemma:pf2}
	For any $w \in \Sigma^{+}$, the components of $\z_{w}^{\vb}(t)$ are orthogonal to $\mathcal{H}_{t+k,+}^{\ub}$, $k \ge 0$.
\end{Lemma}
\begin{pf}[Proof of Lemma \ref{decomp:lemma:pf2}]
	 Let us consider the case $k=0$.
	Since $\mathbf{r}(t):=\begin{bmatrix} \vb(t)^T & \ub(t)^T \end{bmatrix}^T$ is a ZMWSII process , from \cite[Lemma 7]{PetreczkyBilinear} it follows that
	$E[\z_{w}^{\mathbf{r}}(t)(\z^{\mathbf{r}}_{v}(t))^T]=0$ for all $v \in \Sigma^{+}$, $v \ne w$, and 
	if $v=w$ and $\sigma$ is the first letter of $w$, then  $E[\z_{w}^{\mathbf{r}}(t)(\z^{\mathbf{r}}_{w}(t))^T]=E[\z_{\sigma}^{\mathbf{r}}(t)(\z^{\mathbf{r}}_{\sigma}(t))^T]$.
	
	Since $E[\z_{w}^{\vb}(t)(\z^{\ub}_{v}(t))^T]$ is the upper right block of $E[\z_{w}^{\mathbf{r}}(t)(\z^{\mathbf{r}}_{v}(t))^T]$, it follows that 
	$E[\z_{w}^{\vb}(t)(\z^{\ub}_{v}(t))^T]=0$ if $v \ne w$ and  $E[\z_{w}^{\vb}(t)(\z^{\ub}_{w}(t))^T]=E[\z_{\sigma}^{\vb}(t)(\z^{\ub}_{\sigma}(t))^T]=\frac{1}{p_{\sigma}} E[\ub(t-1)\vb(t-1)\p_{\sigma}^2(t-1)]$, and from  Definition \ref{def:Stationary}, it follows that the latter is zero. That is, $E[\z_{w}^{\vb}(t)(\z^{\ub}_{v}(t))^T]=0$ for all $v \in \Sigma^{+}$. 
	
	Finally,  as $\mathbf{r}(t):=\begin{bmatrix} \vb(t)^T & \ub(t)^T \end{bmatrix}^T$ is a 
	white noise process w.r.t. $\p$, it follows that  $E[\z_{w}^{\mathbf{r}}(t)(\mathbf{r}(t))^T]=0$, and since  $E[\z_{w}^{\vb}(t)(\ub(t))^T]$ is the upper right block of $E[\z_{w}^{\mathbf{r}}(t)(\mathbf{r}(t))^T]$, it then follows that $E[\z_{w}^{\vb}(t)(\ub(t))^T]=0$. 
	Since $\z_w^{\vb}(t)$ is orthogonal to the components of the random variables which generate $\mathcal{H}_{t,+}^{\ub}$, the statement of the lemma follows for $k=0$.
	
	Consider now the case $k > 0$. As $\p_1=1$ and $p_1=1$ it follows that $\z_{w}^{\vb}(t)=\z_{w\underline{1\cdots1}_{k}}^{\vb}(t+k)$ (where $\underline{1\cdots1}_{k}$ denotes $k$-lenght word of $1$s), and $\z_{w\underline{1\cdots1}_{k}}^{\vb}(s)$ 
	is orthogonal to $\mathcal{H}_{s,+}^{\ub}$,  according to the case $k=0$. By taking $s=t+k$ for $k > 0$, the statement of the lemma follows. \hfill $\blacksquare$
	\end{pf}

\begin{Lemma}
	\label{decomp:lemma:pf3}
	The components of $\xb^d(t)$ belong to $\mathcal{H}_{t}^{\ub}$ and 
	\begin{equation}
	\label{decomp:lemma:pf3:eq2}
	\xb^d(t) = \sum_{w \in \Sigma^{*}, \sigma \in \Sigma} \sqrt{p_{\sigma w}} A_w B_{\sigma}\z^{\ub}_{\sigma w}(t),
	\end{equation}
	where the right-hand side of \eqref{decomp:lemma:pf3:eq2} converges in the mean square sense. 
\end{Lemma}
\begin{pf}[Proof of Lemma \ref{decomp:lemma:pf3}]
	It is clear from the definition that the components of $\xb^d(t)$ belong to $\mathcal{H}_{t,+}^{\ub}$. Since 
	$\mathbf{x}(t)\! \! = \! \! \sum_{w \in \Sigma^{*}, \sigma \in \Sigma} \sqrt{p_{\sigma w}} A_w \left(K_{\sigma} \z^{\vb}_{\sigma w}(t)  + B_{\sigma}\z^{\ub}_{\sigma w}(t)\right)$
	and the fact that the map $z \mapsto E_l[z \mid M]$ (where $z \in \mathcal{H}_1$) is a continuous linear operator for any closed subspace $M$, it follows that
	\[
	\begin{split}
	 \xb^d(t)=  \sum_{w \in \Sigma^{*}, \sigma \in \Sigma} &\sqrt{p_{\sigma w}} A_w \left(K_{\sigma} E_l[\z^{\vb}_{\sigma w}(t) \mid H_{t,+}^{\ub}] \right. \\ 
	 & \left. + B_{\sigma}E_l[\z^{\ub}_{\sigma w}(t) \mid H_{t,+}^{\ub}]\right).
	\end{split}
	\]
	
	From Lemma \ref{decomp:lemma:pf2} it follows that,  $E_l[\z^{\vb}_{\sigma w}(t) \mid H_{t,+}^{\ub}]=0$, and since the components of $\z^{\ub}_{\sigma w}(t)$ belong to $\mathcal{H}_{t,+}^{\ub}$, it follows that $E_l[\z^{\ub}_{\sigma w}(t) \mid H_{t,+}^{\ub}]=\z^{\ub}_{\sigma w}(t)$.
	
	Hence,
	\begin{equation}
	\label{decomp:lemma:pf3:eq1}
	\begin{split}
	 \xb^d(t) &=\sum_{w \in \Sigma^{*}, \sigma \in \Sigma} \sqrt{p_{\sigma w}} A_w B_{\sigma}E_l[\z^{\ub}_{\sigma w}(t) \mid H_{t,+}^{\ub}] \\
	& = \sum_{w \in \Sigma^{*}, \sigma \in \Sigma} \sqrt{p_{\sigma w}} A_w B_{\sigma}\z^{\ub}_{\sigma w}(t).
	\end{split}
	\end{equation}
	Since the components of $\z^{\ub}_{\sigma w}(t)$ belong to $\mathcal{H}_{t}^{\ub}$, it follows that the components of the right-hand side of 
	\eqref{decomp:lemma:pf3:eq1} belongs to $\mathcal{H}_{t}^{\ub}$ and hence the components of $\xb^d(t)$ belong to $\mathcal{H}_{t}^{\ub}$.
	Note that, the convergence of the right-hand side  of \eqref{decomp:lemma:pf3:eq1} in the mean square sense follows from the convergence of the series $\sum_{w \in \Sigma^{*}, \sigma \in \Sigma} \sqrt{p_{\sigma w}} A_w \left(K_{\sigma} \z^{\vb}_{\sigma w}(t)  + B_{\sigma}\z^{\ub}_{\sigma w}(t)\right)$.
	\hfill $\blacksquare$
	\end{pf}

\begin{Lemma}
	\label{decomp:lemma:pf4}
	The components of $\xb^s(t)$ belong to $\mathcal{H}_{t}^{\vb}$, they are orthogonal to $\mathcal{H}_{t+k,+}^{\ub}$ for any $k \ge 0$ and 
	\begin{equation}
	\label{decomp:lemma:pf4:eq2}
	\xb^s(t) = \sum_{w \in \Sigma^{*}, \sigma \in \Sigma} \sqrt{p_{\sigma w}} A_w K_{\sigma}\z^{\vb}_{\sigma w}(t),
	\end{equation}
	where the right-hand side converges in the mean-square sense.
\end{Lemma}
\begin{pf}[Proof of Lemma \ref{decomp:lemma:pf4}] 
	
	From \eqref{decomp:lemma:pf3:eq2}, $\xb^s(t)=\xb(t)-\xb^d(t)$
	and 
	$\mathbf{x}(t)\! \! = \! \! \sum_{w \in \Sigma^{*}, \sigma \in \Sigma} \sqrt{p_{\sigma w}} A_w \left(K_{\sigma} \z^{\vb}_{\sigma w}(t)  + B_{\sigma}\z^{\ub}_{\sigma w}(t)\right)$, it follows that \eqref{decomp:lemma:pf4:eq2} holds and that its right-hand side converges in the mean square sense. From Lemma \ref{decomp:lemma:pf2}, it follows that for any $w \in \Sigma^{+}$, the components of $\z_{w}^{\vb}(t)$ are orthogonal to $\mathcal{H}_{t+k,+}^{\ub}$, hence
	all the summands of the infinite series of \eqref{decomp:lemma:pf4:eq2} are orthogonal to $\mathcal{H}_{t+k,+}^{\ub}$. \hfill $\blacksquare$
	\end{pf}
Finally, we now state the proof of Lemma  \ref{decomp:lemma} (Decomposition of $\yb$). 
\begin{pf}[Proof of Lemma \ref{decomp:lemma}]
	 It follows that,
	\begin{equation}
	\label{decomp:lemma:pf:eq2}
	\begin{split} 
	& \xb^d(t+1)=E_l[\xb(t+1)  \mid \mathcal{H}_{t+1,+}^{\ub}]  = \\
	& = \sum_{\sigma \in \Sigma} (A_{\sigma} E_l[\xb(t)\p_{\sigma}(t)\mid \mathcal{H}_{t+1,+}^{\ub}] \\
	&+ B_{\sigma} E_l[\ub(t)\p_{\sigma}(t) \mid \mathcal{H}_{t+1,+}^{\ub}] 
	+
	K_{\sigma} E_{l}[\vb(t)\p_{\sigma}(t) \mid \mathcal{H}_{t+1,+}^{\ub}])
	\end{split}
	\end{equation}
	Note that, $\ub(t)\p_{\sigma}(t)=\sqrt{p_{\sigma}} \z^{\ub}_{\sigma}(t+1)$, hence the components of $\ub(t)\p_{\sigma}(t)$ belong to $\mathcal{H}_{t+1,+}^{\ub}$ and
	therefore  
	\[ E_l[\ub(t)\p_{\sigma}(t) \mid \mathcal{H}_{t+1,+}^{\ub}]=\ub(t)\p_{\sigma}(t) \]
	
	We claim that, 
	\[ 
	E_l[\xb(t)\p_{\sigma}(t)\mid \mathcal{H}_{t+1,+}^{\ub}] = \xb^d(t)\p_{\sigma}.
	\]
	Since $\xb(t)=\xb^d(t)+\xb^s(t)$, it follows that $\xb(t)\p_{\sigma}(t)=\xb^d(t)\p_{\sigma}(t)+\xb^s(t)\p_{\sigma}(t)$. 
	
	From \cite[Lemma 9]{PetreczkyBilinear} 
	and Lemma \ref{decomp:lemma:pf3}--\ref{decomp:lemma:pf4}, it follows that
	\begin{equation}
	\label{decomp:lemma:pf:eq3}
	\begin{split}
	& \xb^d(t)\p_{\sigma}(t) = \sum_{w \in \Sigma^{*}, \sigma^{'} \in \Sigma} \sqrt{p_{\sigma^{'} w\sigma}} A_w B_{\sigma^{'}}\z^{\ub}_{\sigma^{'} w\sigma}(t+1) \\
	& \xb^s(t)\p_{\sigma}(t) = \sum_{w \in \Sigma^{*}, \sigma^{'} \in \Sigma} \sqrt{p_{\sigma^{'} w\sigma}} A_w K_{\sigma^{'}}\z^{\vb}_{\sigma^{'} w\sigma}(t+1) \\
	\end{split}
	\end{equation}

From Lemma \ref{decomp:lemma:pf2}, it follows that $\z^{\vb}_{\sigma^{'} w\sigma}(t+1)$ is orthogonal to $\mathcal{H}_{t+1,+}^{\ub}$ for all $w \in \Sigma^{+}$,
$\sigma^{'},\sigma \in \Sigma$, and hence  $\xb_s(t)\p_{\sigma}(t)$ is also orthogonal to $\mathcal{H}_{t+1,+}^{\ub}$.  Moreover, since the components of
$\z^{\ub}_{\sigma^{'} w\sigma}(t+1)$ belong to $\mathcal{H}_{t+1,+}^{\ub}$, it follows that  $\xb^d(t)\p_{\sigma}(t)$ belongs to $\mathcal{H}_{t+1,+}^{\ub}$.
Hence,
\[ 
\begin{split} 
& E_l[\xb(t)\p_{\sigma}(t)\mid \mathcal{H}_{t+1,+}^{\ub}] \\ &=E_l[\xb^s(t)\p_{\sigma}(t) \mid  \mathcal{H}_{t+1,+}^{\ub}] 
+E_l[\xb^d(t)\p_{\sigma}(t) \mid  \mathcal{H}_{t+1,+}^{\ub}]\\
&=E_l[\xb^d(t)\p_{\sigma} \mid  \mathcal{H}_{t+1,+}^{\ub}]=\xb^d(t)\p_{\sigma}(t). 
\end{split}
\]

 Finally, from Lemma \ref{decomp:lemma:pf2}, it follows that $\vb(t)\p_{\sigma}(t)=\sqrt{p_{\sigma}} \z_{\sigma}^{\vb}(t+1)$ is orthogonal to $\mathcal{H}_{t+1,+}^{\ub}$, and hence
\[  E_{l}[\vb(t)\p_{\sigma}(t) \mid \mathcal{H}_{t+1,+}^{\ub}]=0. \]

By collecting all these facts, we can show that 
\[ \begin{split} 
& \xb^d(t+1)=E_l[\xb(t+1)  \mid \mathcal{H}_{t+1,+}^{\ub}]  = \\
& = \sum_{\sigma \in \Sigma} (A_{\sigma} E_l[\xb(t)\p_{\sigma}(t)\mid \mathcal{H}_{t+1,+}^{\ub}] +\\
& B_{\sigma} E_l[\ub(t)\p_{\sigma}(t) \mid \mathcal{H}_{t+1,+}^{\ub}]
+
K_{\sigma} E_{l}[\vb(t)\p_{\sigma}(t) \mid \mathcal{H}_{t+1,+}^{\ub}])\\=
& \sum_{\sigma \in \Sigma} (A_{\sigma}\xb^d(t)+B_{\sigma}\ub(t)) 
\end{split}
\]

That is,  the first equation of \eqref{eqn:LPV_SSA_deter} holds.

As to the second equation of \eqref{eqn:LPV_SSA_deter}, notice that from Definition \ref{decomp:outp},
\[
\begin{split}
&\yb^d(t)=E_l[\yb(t) \mid \mathcal{H}_{t,+}^{\ub}]=\\&=CE_l[\xb(t) \mid \mathcal{H}_{t,+}^{\ub}]+D E_l[\ub(t) \mid \mathcal{H}_{t,+}^{\ub}]+E_l[\vb(t) \mid \mathcal{H}_{t,+}^{\ub}].
\end{split}
\]

Since  the components of $\ub(t)$ are among the generators of $\mathcal{H}_{t,+}^{\ub}$, and by Lemma \ref{decomp:lemma:pf1}, $\vb(t)=\vb(t)\p_1(t)$ is orthogonal to 
$\mathcal{H}_{t,+}^{\ub}$, it follows that $E_l[\vb(t) \mid \mathcal{H}_{t,+}^{\ub}]=0$ and $E_l[\ub(t) \mid \mathcal{H}_{t,+}^{\ub}]=\ub(t)$. 
It then follows that the second equation of \eqref{eqn:LPV_SSA_deter} holds. 

From $\yb^s(t)=\yb(t)-\yb^d(t)$, $\xb^s(t)=\xb(t)-\xb^d(t)$ and \eqref{eqn:LPV_SSA_deter}, \eqref{eqn:LPV_SSA_stoch} follows.

It is left to show that $(\{A_{\sigma},B_{\sigma}\}_{\sigma \in \Sigma},C,D,\xb^d,\ub)$ and $(\{A_{\sigma},K_{\sigma}\}_{\sigma \in \Sigma},C,I_{\ny},\xb^s,\vb)$
are stationary LPV-SSA representations without inputs as per Definition \ref{defn:LPV_SSA_wo_u}. 

Since $\sum_{\sigma \in \Sigma} p_{\sigma} A_{\sigma} \otimes A_{\sigma}$ is stable and $\ub$ and $\vb$ are both white noise processes w.r.t. $\p$, the
only thing which needs to be shown is that $\begin{bmatrix} (\xb^d)^T & \ub^T \end{bmatrix}^T$ and $\begin{bmatrix} (\xb^s)^T & \vb^T \end{bmatrix}^T$
are ZMWSII. However, the latter follows from \eqref{decomp:lemma:pf3:eq2}, \eqref{decomp:lemma:pf4:eq2} and \cite[Lemma 3]{PetreczkyBilinear}. \hfill $\blacksquare$
\end{pf}

\subsection{Proof of Lemma \ref{decomp:lemma:inv}}\label{app:proof_lem2}
We show that $(\{\hat{A}_{\sigma}, \hat{B}_{\sigma}, \hat{K}_{\sigma}\}_{\sigma \in \Sigma},\hat{C},\hat{D},\hat{\xb},\eb^s)$  is a stationary LPV-SSA representation of $\yb$ with input $\ub$, where, $(\{\hat{A}_{\sigma}, \hat{B}_{\sigma}, \hat{K}_{\sigma}\}_{\sigma \in \Sigma},\hat{C},\hat{D},\hat{\xb},\eb^s)$ are as defined in  \eqref{decomp:lemma:sys:eq1}-\eqref{decomp:lemma:innov2}. 

To this end, assume that $(\{A_{\sigma}, B_{\sigma}, K_{\sigma}\}_{\sigma \in \Sigma},C,D,\xb,\vb)$  is a stationary LPV-SSA representation of $\yb$ with input $\ub$.  \footnote{By Assumption of Lemma \ref{decomp:lemma:inv}, such a LPV-SSA representation exists.}
Recall from Notation \ref{hilbert:notation} that $\mathcal{H}_1$ denotes the Hilbert-space of zero mean square integrable random variables.
Denote by $\mathcal{H}_{t,+}^{\vb}$ the closed-subspace of the Hilbert-space $\mathcal{H}_1$ 
generated by the components of $\{\z^{\vb}_w\}_{w \in \Sigma^{+}} \cup \{\vb(t)\}$,   and  denote by
$\mathcal{H}_{t}^{\vb}$ the Hilbert-space generated by $\{\z^{\vb}_w\}_{w \in \Sigma^{+}}$.
We prove the following lemmas. 
\begin{Lemma}\label{decomp:lemma:inv:pf2.1}
  Assume that $(\{A_{\sigma}, B_{\sigma}, K_{\sigma}\}_{\sigma \in \Sigma},C,D,\xb,\vb)$  is a stationary LPV-SSA representation of $\yb$ with input $\ub$ The components of  $\yb^s(t), \z^{\yb^s}_v(t)$, $\eb^s(t)$, $\z^{\eb^s}_v(t)$, $v \in \Sigma^{+}$ belong to $\mathcal{H}_{t,+}^{\vb}$.
\end{Lemma}
\begin{pf}[Proof of Lemma \ref{decomp:lemma:inv:pf2.1}]
Recall from Lemma \ref{decomp:lemma:pf3}, 
		\begin{equation*}
	\xb^s(t) = \sum_{w \in \Sigma^{*}, \sigma \in \Sigma} \sqrt{p_{\sigma w}} A_w K_{\sigma}\z^{\vb}_{\sigma w}(t),
	\end{equation*}
and hence, 
\begin{equation*}
\yb^s(t) =  \sum_{w \in \Sigma^{*}, \sigma \in \Sigma} \sqrt{p_{\sigma w}} C A_w K_{\sigma}\z^{\vb}_{\sigma w}(t)+ \vb(t),
\end{equation*}
That is, the components of $\yb^s(t)$ belong to $\mathcal{H}_{t,+}^{\vb}$. 
In particular, from \cite[Lemma 11]{PetreczkyBilinear}, it follows that the coordinates of
$\z^{\yb^s}_w(t)$ belong to $\mathcal{H}_{t,+}^{\vb}$ and hence, $\mathcal{H}_{t}^{\yb^s} \subseteq \mathcal{H}_{t}^{\vb}$.
Since $\eb^s(t)=\yb^s(t)-E_l[\yb^s(t) \mid \mathcal{H}_{t}^{\yb^s}] $, it follows that the components of 
$\eb^s(t)$ belong to $\mathcal{H}_{t,+}^{\vb}$. Since $\z_{v}^{\vb}(t)=\z_{v1}^{\vb}(t+1)$, $\vb(t)=\z_1^{\vb}(t+1)$, it follows that
 $\mathcal{H}_{t,+}^{\vb} \subseteq \mathcal{H}_{t+1}^{\vb}$ and from  \cite[Lemma 11]{PetreczkyBilinear} it follows that
the components of $\z_v^{\eb^s}(t)$ belong to $\mathcal{H}_{t}^{\vb} \subseteq \mathcal{H}_{t,+}^{\vb}$. 
\end{pf}
\begin{Lemma}
\label{decomp:lemma:inv:pf2.2}
If $\yb$ has a realization by a stationary LPV-SSA representation with input $\ub$, then the components of 
$\yb^s(t), \z^{\yb^s}_v(t), \eb^s(t), \z^{\eb^s}_v(t)$, $v \in \Sigma^{+}$ are orthogonal to $\mathcal{H}_{t,+}^{\ub}$, 
i.e., for all $v,w \in \Sigma^{+}$
\begin{equation}
  \label{decomp:lemma:inv:pf2:eq1}
 \begin{split}
   &E[\eb^s(t)(\z^{\ub}_w(t))^T]=0, ~ 
   E[\ub(t)(\z^{\eb^s}_v(t))^T]=0, ~  \\
   &E[\z_w^{\ub}(t)(\z^{\eb^s}_v(t))^T]=0, \\
&E[\yb^s(t)(\z^{\ub}_w(t))^T]=0, ~ 
   E[\ub(t)(\z^{\yb^s}_v(t))^T]=0, ~  \\
  & E[\z_w^{\ub}(t)(\z^{\yb^s}_v(t))^T]=0 \\
\end{split}
  \end{equation}
\end{Lemma}
\begin{pf}[Lemma \ref{decomp:lemma:inv:pf2.2}]
From Lemma \ref{decomp:lemma:pf1}--\ref{decomp:lemma:pf2} and by noticing that $\vb(t)=\vb(t)\p_1(t)$ it follows that the elements of $\mathcal{H}_{t,+}^{\vb}$ are orthogonal to $\mathcal{H}_{t,+}^{\ub}$.
Hence, the coordinates of $\ub(t)$, $\z_w^{\ub}(t)$, $w \in \Sigma^{+}$ are orthogonal to $\mathcal{H}_{t,+}^{\vb}$. 
Since the coordinates of  $\yb^s(t), \z^{\yb^s}_v(t), \eb^s(t),\z^{\eb^s}_v(t)$ belong to $\mathcal{H}_{t,+}^{\vb}$, it follows 
that the coordinates of 
$\yb^s(t), \z^{\yb^s}_v(t), \eb^s(t),\z^{\eb^s}_v(t)$
are orthogonal to  $\mathcal{H}_{t,+}^{\ub}$. 
Since $\mathcal{H}_{t,+}^{\ub}$ is generated by the coordinates of  $\ub(t)$, $\z_w^{\ub}(t)$, $w \in \Sigma^{+}$, 
\eqref{decomp:lemma:inv:pf2:eq1} follows.
\end{pf}
\begin{Lemma}\label{decomp:lemma:inv:pf2}
	$\begin{bmatrix}  (\eb^s)^T & \ub^T \end{bmatrix}^T$ is a white noise process w.r.t. $\p$ and $E[\eb^s(t)\ub^T(t)\p_{\sigma}^2(t)]=0$ for all $\sigSet$.
\end{Lemma}
\begin{pf}[Proof of Lemma \ref{decomp:lemma:inv:pf2}]
 In order to prove the statement of the lemma, we will first show that $\r(t)=\begin{bmatrix}  (\eb^s)^T & \ub^T \end{bmatrix}^T$ is a ZMWSII, by showing that $\r$  satisfies the conditions of  Definition \ref{def:ZMWSSI} one by one. 
First, we show that the processes $\r(t),\z_w^{\r}(t),w \in \Sigma^{+}$ is zero mean, square integrable.
  
Note that $\ub$ is a white noise process w.r.t. $\p$, in particular,  it is a ZWMSII process and hence $\ub(t),\z_w^{\ub}(t),w \in \Sigma^{+}$ are zero mean, square integrable.
From the fact that $\Sigma_s$ is a stationary LPV-SSA representation of $\yb^s$ it follows that $\eb^s$ is also
a white noise process w.r.t. $\p$, in particular, it is also ZWMSII and thus $\eb^s(t),\z_w^{\eb^s}(t),w \in \Sigma^{+}$ is zero mean, square integrable. From this it follows that  $\r(t)=\begin{bmatrix}  (\eb^s)^T & \ub^T \end{bmatrix}^T$ and
$\z_w^{\r}(t)= \begin{bmatrix}  (\z^{\eb^s}_w(t))^T & (\z^{\ub}_w(t))^T \end{bmatrix}^T$ are zero mean and square integrable.

 From Lemma \ref{decomp:lemma:inv:pf2.1} it follows that $\eb^s(t)$ belongs to $\mathcal{H}_{t,+}^{\vb}(t)$, where $\vb$ is a noise process of a stationary LPV-SSA representation of
 $\yb$ with input $\ub$. From the definition of a stationary LPV-SSA representation it then follows that $\begin{bmatrix} \vb^T & \ub^T \end{bmatrix}^T$ is ZMWSII. 
 Hence, with the notation of 
 Definition \ref{def:ZMWSSI}, the $\sigma$-algebras $\mathcal{F}_t^{\begin{bmatrix} \vb^T & \ub^T \end{bmatrix}^T}$ and $\mathcal{F}_t^{\p,+}$ are conditionally independent  w.r.t. $\mathcal{F}_t^{\p,-}$. 
 From the fact that $\eb^s(t)$ belongs to $\mathcal{H}_{t,+}^{\vb}(t)$ it follows that $\eb^s(t)$ is measurable with respect to the $\sigma$-algebra generated by
 $\{\vb(t)\} \cup \{ \z_v^{\vb}(t) \mid v \in \Sigma^{+}\}$ and the latter $\sigma$-algebra is a subset of $\mathcal{F}^{\begin{bmatrix} \vb^T & \ub^T \end{bmatrix}^T}_t \lor \mathcal{F}^{\p,-}_t$, 
 where for two $\sigma$-algebras $\mathcal{F}_i$, $i=1,2$, $\mathcal{F}_1 \lor \mathcal{F}_2$ denotes the smallest $\sigma$-algebra generated by the $\sigma$-algebras $\mathcal{F}_1,\mathcal{F}_2$. 
  That is, $\eb^s(t)$ is measurable w.r.t. the $\sigma$ algebra $\mathcal{F}^{\begin{bmatrix} \vb^T & \ub^T \end{bmatrix}^T}_t \lor \mathcal{F}^{\p,-}_t$
  $\mathcal{F}_t^{\begin{bmatrix} (\eb^s)^T & \ub^T \end{bmatrix}^T} \subseteq \mathcal{F}^{\begin{bmatrix} \vb^T & \ub^T \end{bmatrix}^T}_t \lor \mathcal{F}^{\p,-}_t$. Since  $\mathscr{F}^{\begin{bmatrix} \vb^T & \ub^T \end{bmatrix}^T}_t$ and $\mathscr{F}^{\p,+}$ are conditionally independent w.r.t.  $\mathscr{F}^{\p,-}_t$, from \cite[Proposition 2.4]{vanputten1985} it follows that
  $\mathscr{F}^{\begin{bmatrix} \vb^T & \ub^T \end{bmatrix}^T}_t \lor \mathscr{F}^{\p,-}_t$ and $\mathscr{F}^{\p,+}_t$ are conditionally independent w.r.t. 
 $\mathscr{F}^{\p,-}_t$, and as $\mathscr{F}^{\begin{bmatrix} (\eb^s)^T & \ub^T \end{bmatrix}^T}_t \subseteq \mathscr{F}^{\begin{bmatrix} \vb^T & \ub^T \end{bmatrix}^T}_t \lor \mathscr{F}^{\p,-}_t$, 
 it follows that   $\mathscr{F}^{\begin{bmatrix} (\eb^s)^T & \ub^T \end{bmatrix}^T}_t$ and $\mathscr{F}^{\p,+}$ are conditionally independent w.r.t. $\mathscr{F}^{\p,-}_t$.

 Finally,  from \eqref{decomp:lemma:inv:pf2:eq1} it follows that $\r(t),\z_w^{\r}(t),w \in \Sigma^{+}$, $\r(t)=\begin{bmatrix}  (\eb^s)^T & \ub^T \end{bmatrix}^T$
 are   jointly wide-sense stationary, i.e., for all $s,t \in \mathbb{Z}$, $s \le t$, $v,w \in  \Sigma^{+}$, 
  \begin{align*}
		\expect{\r(t+k)(\zwr(s+k))^{T}} &= \expect{\r(t) (\zwr(s))^{T}}, \\
		\expect{\r(t+k)(\r(s+k))^{T}} &= \expect{\r(t)(\r(s))^{T}}, \\
		\expect{\zwr(t+k)(\zvr(s+k))^{T}} &= \expect{\zwr(t) (\zvr(s))^{T}}.
 \end{align*}
 Indeed, from  \eqref{decomp:lemma:inv:pf2:eq1} and the fact that $\eb_s$, $\ub$ are ZMWSII and hence  the processes $\eb^s(t),\z_w^{\eb^s}(t),w \in \Sigma^{+}$ are jointly wide-sense stationary
 and the processes $\ub(t),\z_w^{\ub}(t),w \in \Sigma^{+}$ are also jointly wide-sense stationary, 
it follows that 
  \begin{align*}
	&	\expect{\r(t+k)(\zwr(s+k))^{T}} =\\
       &\begin{bmatrix} \expect{\eb_s(t+k)(\z_{w}^{\eb_s}(s+k)} & 0 \\ 0 & \expect{\ub(t+k)(\z_{w}^{\ub}(s+k)} \end{bmatrix} = \\
        &     \begin{bmatrix} \expect{\eb^s(t)(\z_{w}^{\eb^s}(s)} & 0 \\ 0 & \expect{\ub(t)(\z_{w}^{\ub}(s)} \end{bmatrix} =   \expect{\r(t) (\zwr(s))^{T}}, \\
	&	\expect{\r(t+k)(\r(s+k))^{T}} = \\
  & \begin{bmatrix} \expect{\eb^s(t+k)(\eb^s(s+k))^T} & 0 \\ 0 & \expect{\ub(t+k)(\ub(s+k))^T} \end{bmatrix} = \\
        &     \begin{bmatrix} \expect{\eb^s(t)(\eb^s(s))^T} & 0 \\ 0 & \expect{\ub(t)(\ub(s))^T} \end{bmatrix} = \expect{\r(t)(\r(s))^{T}}, \\
		& \expect{\zwr(t+k)(\zvr(s+k))^{T}} = \\
    & \begin{bmatrix} \expect{\z^{\eb^s}_w(t+k)(\z_{v}^{\eb^s}(s+k))^T} & 0 \\ 0 & \expect{\z^{\ub}_w(t+k)(\z_{v}^{\ub}(s+k)} \end{bmatrix} = \\
        &     \begin{bmatrix} \expect{\z^{\eb^s}_w(t)(\z_{v}^{\eb^s}(s))^T} & 0 \\ 0 & \expect{\z^{\ub}_w(t)(\z_{v}^{\ub}(s))^T} \end{bmatrix}\\& =  \expect{\zwr(t) (\zvr(s))^{T}}.
 \end{align*}
   Above, we used the fact that if $s < t$, then $\ub(s+k)=\z_{h}^{\ub}(t+k+1)$, $\eb^s(s+k)=\z_{h}^{\eb^s}(t+k+1)$, $\ub(s)=\z_{h}^{\ub}(t+1)$, $\eb^s(s)=\z_{h}^{\eb^s}(t+1)$,
   $\z^{\ub}_w(s+k)=\z_{wh}^{\ub}(t+k)$, $\z_{v}^{\eb^s}(s+k)=\z_{vh}^{\eb^s}(t+k)$, where $h=\underbrace{1\cdots 1}_{t-s}$. 

 That is, we have shown that $\r(t)=\begin{bmatrix}  (\eb^s)^T & \ub^T \end{bmatrix}^T$ satisfies all the conditions of  Definition \ref{def:ZMWSSI}.

Next we show that $\r(t)$ is a white noise process w.r.t. $\p$, i.e., $E[\r(t)(\z^{\r}_w(t))^T]=0$ for all $w \in \Sigma^{+}$. From 
\eqref{decomp:lemma:inv:pf2:eq1} it follows that 
\[
   \expect{\r(t) (\zwr(t))^{T}}= 
   \begin{bmatrix} \expect{\eb^s(t)(\z_{w}^{\eb^s}(t)} & 0 \\ 0 & \expect{\ub(t)(\z_{w}^{\ub}(t)} \end{bmatrix} 
\]
 Notice $\eb_s$ is a white noise process w.r.t. $\p$, since $\Sigma_s=(\{\hat{A}^{s}_{i}, \hat{K}^{s}_{i} \}_{i=1}^{\pdim}, \hat{C}^{s}, I_{\ny},\hat{\xb}^s,\eb^{s})$ is a stationary
 LPV-SSA representation of $\yb^s$ without inputs, and hence  $\expect{\eb_s(t)(\z_{w}^{\eb_s}(t)}=0$. 
 Furthermore, $\ub$ is a white noise process w.r.t. $\p$ by assumption, so $\expect{\ub(t)(\z_{w}^{\ub}(t)}=0$. Hence, $\expect{\r(t) (\zwr(t))^{T}}=0$.

 It is left to show that $E[\eb^s(t)\ub^T(t)\p_{\sigma}^2(t)]=0$ for all $\sigSet$. Notice that $E[\eb^s(t)\ub^T(t)\p_{\sigma}^2(t)]=E[\z^{\eb^s}_{\sigma}(t+1)(\z^{\ub}_{\sigma}(t+1))^T]p_{\sigma}$ by definition, and from \eqref{decomp:lemma:inv:pf2:eq1} it follows that $E[\z^{\eb^s}_{\sigma}(t+1)(\z^{\ub}_{\sigma}(t+1))^T]=0$.
\end{pf}
\begin{Lemma}\label{decomp:lemma:inv:pf3}
	$\sum_{\sigma \in \Sigma} p_{\sigma} \hat{A}_{\sigma} \otimes \hat{A}_{\sigma}$ is stable, where $\hat{A}_{\sigma}= \mathrm{diag}(\hat{A}^{d}_{\sigma},\hat{A}^{s}_{\sigma})$. 
\end{Lemma}
\begin{pf}[Proof of Lemma \ref{decomp:lemma:inv:pf3}]
	From \cite[Proposition 2.6]{CostaBook}, it follows that $\sum_{\sigma \in \Sigma} p_{\sigma} \hat{A}^d_{\sigma} \otimes \hat{A}^d_{\sigma}$ and $\sum_{\sigma \in \Sigma} p_{\sigma} \hat{A}^s_{\sigma} \otimes \hat{A}^s_{\sigma}$, are stable if $\exists \ Q^d, Q^s > 0$ such that,
	\[
	\begin{split}
& Q^d - \sum_{\sigma \in \Sigma} p_{\sigma} \hat{A}^d_{\sigma} Q^d  (\hat{A}^d_{\sigma})^T >0  \\
& Q^s - \sum_{\sigma \in \Sigma} p_{\sigma} \hat{A}^s_{\sigma} Q^s  (\hat{A}^s_{\sigma})^T >0 
	\end{split}
	\]
	
It then follows that,
\begin{equation*}
 \begin{bmatrix}
Q^d - \sum_{\sigma \in \Sigma} p_{\sigma} \hat{A}^d_{\sigma} Q^d  (\hat{A}^d_{\sigma})^T &0 \\ 
0  & Q^s - \sum_{\sigma \in \Sigma} p_{\sigma} \hat{A}^s_{\sigma} Q^s  (\hat{A}^s_{\sigma})^T
\end{bmatrix} > 0
\end{equation*}

\begin{equation*}
=\underset{Q>0}{\underbrace{\begin{bmatrix}
Q^d  &0 \\ 
0  & Q^s 
\end{bmatrix}}} - 
\sum_{\sigma \in \Sigma} p_{\sigma} 
\begin{bmatrix}
 \hat{A}^d_{\sigma} &0 \\ 
0  &  \hat{A}^s_{\sigma} 
\end{bmatrix}
\underset{Q>0}{ \underbrace{
\begin{bmatrix}
Q^d&0 \\ 
0  &  Q^s
\end{bmatrix} }}
\begin{bmatrix}
\hat{A}^d_{\sigma} &0 \\ 
0  &  \hat{A}^s_{\sigma} 
\end{bmatrix}^T > 0
\end{equation*}

From the corollary, \cite[Proposition 2.6]{CostaBook}, it follows that, $\sum_{\sigma \in \Sigma} p_{\sigma} \hat{A}_{\sigma} \otimes \hat{A}_{\sigma}$ is stable, with $\hat{A}_{\sigma}= \mathrm{diag}(\hat{A}^{d}_{\sigma},\hat{A}^{s}_{\sigma})$. \hfill $\blacksquare$
	
	\end{pf}
\begin{Lemma}
	\label{decomp:lemma:inv:pf1}
	 The processes $\hat{\xb}, \ub, \eb$ satisfy all the conditions for noise and state processes of a stationary LPV-SSA representation with no inputs, i.e., 
  $\begin{bmatrix} \hat{\xb}^T & \ub^T & (\eb^s)^T \end{bmatrix}^T$
is a ZMWSII, 
 $\begin{bmatrix}  (\eb^s)^T & \ub^T \end{bmatrix}^T$ is a white noise process w.r.t. $\p$, and 
 $E[\hat{\xb}(t)(\z^{[\ub \ \eb^s]}_w(t))^T] = 0$ and $E[\z_{\sigma}^{\hat{\xb}}(t)(\z^{[\ub \ \eb^s]}_\sigma(t))^T] = 0$.
\end{Lemma}

\begin{pf}[Proof of Lemma \ref{decomp:lemma:inv:pf1}]
       It follows from Lemma \ref{decomp:lemma:inv:pf2} that $\begin{bmatrix}  (\eb^s)^T & \ub^T \end{bmatrix}^T$ is a white noise process w.r.t. $\p$.

	From \cite[Lemma 2]{PetreczkyBilinear}, it follows that 
	
	\begin{equation}
	\label{decomp:lemma:inv:pf1:eq1}
	\hat{\xb}^d(t) = \sum_{w \in \Sigma^{*}, \sigma \in \Sigma} \sqrt{p_{\sigma w}} \hat{A}^d_w \hat{B}_{\sigma}\z^{\ub}_{\sigma w}(t),
	\end{equation}
	
	and  
	\begin{equation}
	\label{decomp:lemma:inv:pf1:eq2}
	\hat{\xb}^s(t) = \sum_{w \in \Sigma^{*}, \sigma \in \Sigma} \sqrt{p_{\sigma w}} \hat{A}^s_w \hat{K}^s_{\sigma}\z^{\eb^s}_{\sigma w}(t),
	\end{equation}
	
Thus, 
\begin{equation}
\label{decomp:inv:pf3:eq1}
\begin{split}
&\hat{\xb}(t) = \begin{bmatrix} (\hat{\xb}^d(t))^T  & (\hat{\xb}^s(t))^T \end{bmatrix}^T= \\
&
 = \sum_{w \in \Sigma^{*}, \sigma \in \Sigma} \sqrt{p_{\sigma w}} \begin{bmatrix}
 \hat{A}^d_{w} &0 \\ 
 0  &  \hat{A}^s_{w} 
 \end{bmatrix} \begin{bmatrix}
 \hat{B}^d_{\sigma} & 0 \\ 
  0& \hat{K}^s_{\sigma} 
 \end{bmatrix} \begin{bmatrix}
\z^{\ub}_{\sigma w}(t) \\ 
\z^{\eb^s}_{\sigma w}(t)
 \end{bmatrix} \\
 & = \sum_{w \in \Sigma^{*}, \sigma \in \Sigma} \sqrt{p_{\sigma w}}
\hat{A}_{w}  
\begin{bmatrix}
 \hat{B}_{\sigma} &\hat{K}_{\sigma}   
  \end{bmatrix}
  \begin{bmatrix}
  \z^{\ub}_{\sigma w}(t) \\ 
  \z^{\eb^s}_{\sigma w}(t)
  \end{bmatrix}, 
\end{split}.
\end{equation}
From Lemma \ref{decomp:lemma:inv:pf3}, it follows that	$\sum_{\sigma \in \Sigma} p_{\sigma} \hat{A}_{\sigma} \otimes \hat{A}_{\sigma}$ is stable.
That is, the noise process $\begin{bmatrix}  (\eb^s)^T & \ub^T \end{bmatrix}^T$ and the matrices $\{\hat{A}_{\sigma}\}_{\sigma \in \Sigma}$ satisfy the conditions of a stationary
LPV-SSA representation without inputs. From \cite[Lemma 3]{PetreczkyBilinear} it then  follows that $\hat{\xb}$ is the unique process such that
$\begin{bmatrix} \ub^T & (\eb^s)^T \end{bmatrix}^T$ satisfies all the conditions of a stationary LPV-SSA representation. 
\end{pf}
\begin{pf}[Proof of Lemma \ref{decomp:lemma:inv}]
It is clear from Lemmas \ref{decomp:lemma:inv:pf1}, \ref{decomp:lemma:inv:pf2}, \ref{decomp:lemma:inv:pf3} that 
$\{A_{\sigma}\}_{\sigma \in \Sigma}$ satisfies the conditions of the definition of a stationary LPV-SSA representation without inputs. From Lemma \ref{decomp:lemma:inv:pf2}, it follows that $\hat{\xb}$ and $\begin{bmatrix} (\eb^s)^T  & \ub^T \end{bmatrix}^T$ satisfies the conditions of an LPV-SSA representation without inputs.
From Lemmas \ref{decomp:lemma:inv:pf1} it follows that the noise process $\eb^s$ and the input $\ub$ satisfy the condition of $E[\eb^s(t)(\ub(t))^T\p_{\sigma}^2(t)]=0$, $\sigma \in \Sigma$.
Hence, $(\{\hat{A}_{\sigma}, \hat{B}_{\sigma}, \hat{K}_{\sigma}\}_{\sigma \in \Sigma},\hat{C},\hat{D},\hat{\xb},\eb^s)$  is a stationary LPV-SSA representation of $\yb$ with input $\ub$. 
\end{pf}

\subsection{Proof of Lemma \ref{thm:cra}}
\label{app:proof_lem4}

\begin{pf}[Proof of Lemma \ref{thm:cra}]

\textbf{If $(\{A_{\sigma},B_{\sigma}\}_{\sigma \in \Sigma},C,D,\xb,\ub)$ is a stationary LPV-SSA representation without inputs of $\yb^d$ $\implies$ $\incov=M_{\mathscr{S}}$, where $\mathscr{S}=(\{A_{\sigma},B_{\sigma}\}_{\sigma \in \Sigma},C,D)$. }

Recall that,
\begin{equation*}
	\yb^d(t) =  \sum_{s \in \Sigma^{*}, \sigma \in \Sigma} \sqrt{p_{\sigma w}} C A_s B_{\sigma}\z^{\ub}_{\sigma s}(t)+ D \ub(t)
	\end{equation*}
Consider, 
\[ 
\begin{split}
&E[\yb^d(t) (\z_w^{\ub}(t))^T] = \\
&\sum_{s \in \Sigma^{*}, \sigma \in \Sigma} \sqrt{p_{\sigma s}} C A_s B_{\sigma}E[\z^{\ub}_{\sigma s}(t) (\z_w^{\ub}(t))^T] \!\! +\!\! 
DE[ \ub(t) (\z_w^{\ub}(t))^T] \\
&= \sqrt{p_{\sigma s}} C A_s B_{\sigma} \Lambda_{\ub}
\end{split}
\]
This follows, as  $\ub$ is white noise process, $E[ \ub(t) (\z_w^{\ub}(t))^T] =0$ and $E[\z^{\ub}_{\sigma s}(t) (\z_w^{\ub}(t))^T] = \Lambda_{\ub}$ if $\sigma s = w$, otherwise $E[\z^{\ub}_{\sigma s}(t) (\z_w^{\ub}(t))^T] = 0$.

Similarly,  $E[\yb^d(t)\ub(t)]= D \Lambda_{\ub}$.

Finally, we recall that $\yb(t) = \yb^d(t) + \yb^s(t)$.
From Lemma \ref{decomp:lemma:inv:pf2.2} it follows that
that the he components of $\yb^s(t)$ is orthogonal to $\mathcal{H}^{\ub}_{t,+}$, that is,
$E[\yb^s(t) (\z^{\ub}_w(t))^T] = 0$ and  $E[\yb^s(t) (\ub(t))^T] = 0$. 
Thus, we have, $E[\yb(t) (\z^{\ub}_w(t))^T] = E[\yb^d(t)(\z^{\ub}_w(t))^T]$, $E[\yb(t) (\ub(t))^T] = E[\yb^d(t)(\ub(t))^T]$ and the statement of Lemma follows.

\textbf{$\hat{\mathscr{S}}=(\{\hat{A}_{\sigma},\hat{B}_{\sigma}\}_{\sigma \in \Sigma},\hat{C},\hat{D})$ is a minimal dimensional determinstic LPV-SSA representation such that $M_{\hat{\mathscr{S}}}=\incov$ $\implies$ 
$(\{\hat{A}_{\sigma},\hat{B}_{\sigma}\},\hat{C},\hat{D},\hat{\xb},\ub)$ is a stationary LPV-SSA representation (without inputs) of $\yb^d$}.
Consider the formal power series \citep[Appendix C]{PetreczkyBilinear}
of $\Psi(w) = \left[ M_{\hat{\mathscr{S}}}(1w)\sqrt{p_{1w}} \ \ \  M_{\hat{\mathscr{S}}}(\pdim w)\sqrt{p_{{\pdim}w}} \right]$.
Let $\tilde{\mathscr{S}}=(\{\tilde{A}_{\sigma},\tilde{B}_{\sigma}\}_{\sigma \in \Sigma},\tilde{C},\tilde{D})$ be any deterministic LPV-SSA representation such that $M_{\tilde{\mathscr{S}}}$. 
Consider the recognizable representation in the sense of \citep[Appendix C]{PetreczkyBilinear} defined as
$R_{\tilde{\mathscr{S}}} =(\{\sqrt{p_{\sigma}} \tilde{A}_{\sigma}\}_{\sigma \in \Sigma}, \tilde{B}, \tilde{C})$, $\tilde{B} = \left[ \sqrt{p_1} \tilde{B}_{1} \ \cdots \ \sqrt{p_{\pdim}} \tilde{B}_{\pdim}\right]$. We claim that $R_{\tilde{\mathscr{S}}}$ is a recognizable representation of $\Psi$ (see
\citep[Appendix C]{PetreczkyBilinear} for the definition of a recognizable representation of a formal power series),
and if $\tilde{\mathscr{S}}$ is a minimal dimensional deterministic LPV-SSA representation such that
$M_{\tilde{\mathscr{S}}}=\incov$, then $R_{\tilde{\mathscr{S}}}$ is a minimal dimensional representation of $\Psi$.
For the definition of the dimension of a recognizable representation, see \citep[Appendix C]{PetreczkyBilinear}.
We call $R_{\tilde{\mathscr{S}}}$ \emph{the recognizable representation associated with the deterministic LPV-SSA representation
$\mathscr{S}$}.

Indeed, $M_{\tilde{\mathscr{S}}}(\sigma w) \sqrt{p_{\sigma}w} =  \sqrt{p_w}\tilde{C}\tilde{A}_w\tilde{B}_{\sigma} \sqrt{p_{\sigma}}$, hence $\Psi(w)=\tilde{C}F_w\tilde{B}$, $F_{\sigma}=\sqrt{p_{\sigma}}\tilde{A}_{\sigma}$, 
which by definition \cite[Appendix C]{PetreczkyBilinear}
means that $R$ is a representation of $\Psi$. 

In order to show that $R_{\tilde{\mathscr{S}}}$ is a minimal representation of $\Psi$, if $\tilde{\mathscr{S}}$ is a 
minimal dimensional deterministic LPV-SSA representation such that 
$M_{\tilde{\mathscr{S}}}=\incov$, we proceed as follows.  
Consider a recognizable representation $R_{\mathrm{o}}= (\{\bar{F}_{\sigma} \}_{\sigSet}, \bar{G}, \bar{C} )$ of $\Psi$.
Define the deterministic LPV-SSA representation
$\bar{\mathscr{S}}=(\{\{\bar{A}_{\sigma},\bar{B}_{\sigma}\}_{\sigma \in \Sigma},\bar{C},D)$, where $\bar{A}_{\sigma}=\frac{1}{\sqrt{p_{\sigma}}} \bar{F}_{\sigma}$, and $\bar{G}=\left[ \sqrt{p_1} \tilde{B}_{1} \ \cdots \ \sqrt{p_{\pdim}} \tilde{B}_{\pdim}\right]$. Then, since $\Psi(w)=\bar{C}\bar{F}_w\bar{G}$, $w \in \Sigma^{*}$, it follows that
$M_{\tilde{\mathscr{S}}}(\sigma w)=\frac{1}{\sqrt{p_w}\sqrt{p_{\sigma}}} \bar{C}\bar{F}_w (\sqrt{p_{\sigma}} \tilde{B}_{\sigma})=\bar{C}\bar{A}_w\bar{B}_{\sigma}$, i.e., $M_{\bar{\mathscr{S}}}=M_{\tilde{\mathscr{S}}}=\incov$. By the assumption,
$\tilde{\mathscr{S}}$ is a minimal dimensional determinsitic LPV-SSA representation such that
$M_{\tilde{\mathscr{S}}}=\incov$, hence the dimension of $\bar{\mathscr{S}}$ should not be smaller than that of $\tilde{\mathscr{S}}$. However, the dimension of $\tilde{\mathscr{S}}$ equals the dimension of the representation $R_{\tilde{\mathscr{S}}}$, and the
dimension of $\bar{\mathscr{S}}$ equals the dimension of $R_{\mathrm{o}}$. That is, the dimension of any representation of $\Psi$ cannot be smaller than the dimension of $R_{\tilde{\mathscr{S}}}$, i.e., $R_{\tilde{\mathscr{S}}}$ is minimal.

Since it is assumed that $\yb$ has a stationary LPV-SSA representation with input $\ub$, from Lemma \ref{decomp:lemma}
it follows that there exists a stationary LPV-SSA representation 
$(\{A_{\sigma},B_{\sigma}\},C,D,\xb^d,\ub)$ of $\yb^d$ without inputs. Then by the first part of
this lemma, the deterministic LPV-SSA representation $\mathscr{S}=(\{A_{\sigma},B_{\sigma}\},C,D)$ is such that
$M_{\mathscr{S}}=\incov$. Since by definition of a stationary LPV-SSA representation without inputs, 
$\sum_{\sigma \in \Sigma} p_{\sigma} A_{\sigma} \otimes A_{\sigma}=\sum_{\sigma \in \Sigma} (\sqrt{p_{\sigma}} A_{\sigma}) \otimes (\sqrt{p_{\sigma}} A_{\sigma})$ is stable, then
using the terminology of \cite[Appendix C]{PetreczkyBilinear}, 
 the representation $R_{\mathscr{S}}$ associated with $\mathscr{S}$
is  a stable representation. Recall from \cite[Appendix C]{PetreczkyBilinear} that a 
recognizable representation $R=(\{F_{\sigma}\}_{\sigma \in \Sigma},G,H)$ is stable, if all eigenvalues of the matrix
$\sum_{\sigma \in \Sigma} F_{\sigma} \otimes F_{\sigma}$ are inside the open  unit disk. 
Since by the discussion above $R_{\mathscr{S}}$ is a representation of $\Psi$, 
by \cite[Theorem 6]{PetreczkyBilinear} $\Psi$ is square summable (see \cite[Appendix C]{PetreczkyBilinear} for the definition of a square summable formal power series).

Consider now the minimal deterministic LPV-SSA representation $\hat{\mathscr{S}}$ from the statement of the lemma.
From the discussion above it then follows that the associated recognizable representation $R_{\hat{\mathscr{S}}}=(\{\sqrt{p_{\sigma}} \hat{A}_{\sigma} \}_{\sigma \in \Sigma},B,C)$
is a minimal representation of $\Psi$. Since $\Psi$ is square summable and $R_{\hat{\mathscr{S}}}$ is minimal, 
from \cite[Theorem 6]{PetreczkyBilinear} it follows that $R_{\hat{\mathscr{S}}}$ is stable, which means that all
the eigenvalues of 
$\sum_{\sigma \in \Sigma} p_{\sigma} \hat{A}_{\sigma} \otimes \hat{A}_{\sigma}$ are inside the open unit disk. 
Notice that $\ub$ is a white noise process w.r.t. $\p$ by assumption. Then it follows from 
\cite[Lemma 3]{PetreczkyBilinear} that with 
$\hat{\xb}(t)=\sum_{w \in \Sigma^{*},\sigma \in \Sigma} \sqrt{p_{\sigma w}} \hat{A}_w\hat{B}_{\sigma}\z^{\ub}_{\sigma w}(t)$
$(\{\hat{A}_{\sigma},\hat{B}_{\sigma}\},\hat{C},\hat{D},\hat{\xb},\ub)$ is a stationary LPV-SSA representation. 

It is left to show that $(\{\hat{A}_{\sigma},\hat{B}_{\sigma}\},\hat{C},\hat{D},\hat{\xb},\ub)$ is 
a representation of $\yb^d$. Consider  the representation stationary LPV-SSA representation 
$(\{A_{\sigma},B_{\sigma}\},C,D,\xb^d,\ub)$ of $\yb^d$ without inputs from the discussion above. 
It then follows from \cite[Lemma 1]{PetreczkyBilinear} that 
\begin{align*}
 & \yb^d(t)=\sum_{w \in \Sigma^{*},\sigma \in \Sigma} \sqrt{p_{\sigma w}} \sqrt{p_{\sigma w}} CA_wB_{\sigma}\z^{\ub}_{\sigma w}(t) + D\ub(t) \\
 & = \sum_{w \in \Sigma^{*},\sigma \in \Sigma} \sqrt{p_{\sigma w}} \underbrace{M_{\mathscr{S}}(\sigma w)}_{=\incov(\sigma w)=M_{\hat{\mathscr{S}}}(\sigma w)=\hat{C}\hat{A}_w\hat{B}_{\sigma}}  \z^{\ub}_{\sigma w}(t)+ \\
& \underbrace{M_{\mathscr{S}}(\epsilon)}_{\incov(\epsilon)=M_{\hat{\mathscr{S}}}(\epsilon)=\hat{D}}\ub(t) \\
 & = \sum_{w \in \Sigma^{*},\sigma \in \Sigma} \sqrt{p_{\sigma w}}  \hat{C}\hat{A}_w \hat{B}_{\sigma} \z^{\ub}_{\sigma w}(t)+\hat{D}\ub(t) = \hat{C}\hat{\xb}(t)+\hat{D}\ub(t) \\
\end{align*}
where we used that by \cite[Lemma 3]{PetreczkyBilinear} 
$\hat{\xb}(t)=\sum_{w \in \Sigma^{*},\sigma \in \Sigma} \sqrt{p_{\sigma w}} \hat{A}_w\hat{B}_{\sigma}\z^{\ub}_{\sigma w}(t)$. This means that $\yb^d(t)=\hat{C}\hat{\xb}(t)+\hat{D}\ub(t)$, and hence $(\{\hat{A}_{\sigma},\hat{B}_{\sigma}\},\hat{C},\hat{D},\hat{\xb},\ub)$ is  a representation of $\yb^d$.

\end{pf}

\subsection{Proof of Lemma \ref{lemma:stoch-real1}}\label{app:proof_lem5}

Assume $\mathscr{S}=(\{\hat{A}_{\sigma},\hat{\Bs}_{\sigma}\}_{\sigma \in \Sigma},\hat{C},I_{\ny})$ is  LPV-SSA representation whose sub-Markov parameters are $M_{\mathscr{S}}=\covseq$. 

Let $\Psi(w)= \left[\covseq(1w) \cdots \covseq(\pdim w) \right]$, $\forall \wordSet{*}$. Then, $R = (\{\hat{A}_{\sigma},\}_{\sigma \in \Sigma}, \hat{\Bs}_{}, \hat{C})$, with $\hat{\Bs}= \left[\hat{\Bs}_{1} \cdots \hat{\Bs}_{\pdim} \right]$, is a representation of $\Psi$ and by \cite[Theorem 1, Theorem 2]{PetreczkyLPVSS} and the definition of
observability and reachability for representations in \cite[Appendix C]{PetreczkyBilinear}, it follows that
 $R$ is observable and reachable, and hence $R$ is minimal by \cite{Reut:Book,Son:Real}. 

Then, the statement follows from \cite[Theorem 5 and Lemma 21]{PetreczkyBilinear}. \hfill $\blacksquare$.  
\end{document}